\documentclass[journal]{IEEEtran}
\usepackage[table]{xcolor}
\usepackage{multirow}
\usepackage{mathtools}
\usepackage{cite}
\usepackage{textcomp}
\usepackage{tabu}
\usepackage{booktabs}
\usepackage{hyperref}
\usepackage{amsmath,amssymb,amsfonts}
\usepackage{algorithm}
\usepackage{graphicx}
\usepackage{xspace}
\usepackage{balance}
\usepackage[inline]{enumitem}
\usepackage{amsmath}
\usepackage[noend]{algpseudocode}
\usepackage{subcaption} 
\usepackage{url} 
\usepackage{stfloats}
\usepackage[colorinlistoftodos,prependcaption,textsize=tiny]{todonotes}
\usepackage{cleveref}
\usepackage[bottom]{footmisc}
\usepackage{siunitx}
\usepackage{array}
\newcolumntype{;}{>{\global\let\currentrowstyle\relax}}
\newcolumntype{^}{>{\currentrowstyle}}

\usepackage{microtype} 
\crefformat{section}{\S#2#1#3}
\crefformat{subsection}{\S#2#1#3}
\crefformat{subsubsection}{\S#2#1#3}

\SetLabelAlign{LeftAlignWithIndent}{\hspace*{-0.9ex}\makebox[1.15em][l]{#1}}

\DeclareMathOperator*{\argmax}{arg\,max}
\newcommand{\sysname}{DRLE\xspace}

\newcommand\CONDITION[2]%
{\begin{tabular}[t]{@{}l@{}l@{}}
		#1&#2
	\end{tabular}%
}

\algdef{SE}[WHILE]{While}{EndWhile}[1]%
{\algorithmicwhile\ \CONDITION{#1}{\ \algorithmicdo}}%
{\algorithmicend\ \algorithmicwhile}
\algdef{SE}[FOR]{For}{EndFor}[1]%
{\algorithmicfor\ \CONDITION{#1}{\ \algorithmicdo}}%
{\algorithmicend\ \algorithmicfor}
\algdef{S}[FOR]{ForAll}[1]%
{\algorithmicforall\ \CONDITION{#1}{\ \algorithmicdo}}
\algdef{SE}[REPEAT]{Repeat}{Until}{\algorithmicrepeat}[1]%
{\algorithmicuntil\ \CONDITION{#1}{}}
\algdef{SE}[IF]{If}{EndIf}[1]%
{\algorithmicif\ \CONDITION{#1}{\ \algorithmicthen}}%
{\algorithmicend\ \algorithmicif}%
\algdef{C}[IF]{IF}{ElsIf}[1]%
{\algorithmicelse\ \algorithmicif\ \CONDITION{#1}{\ \algorithmicthen}}
\def\BibTeX{{\rm B\kern-.05em{\sc i\kern-.025em b}\kern-.08em
    T\kern-.1667em\lower.7ex\hbox{E}\kern-.125emX}}

\begin{document}

\title{\sysname: Decentralized Reinforcement Learning at the Edge for Traffic
Light Control in the IoV}

\author{Pengyuan Zhou,~\IEEEmembership{Member,~IEEE},
Xianfu Chen,~\IEEEmembership{Member,~IEEE},
Zhi Liu,~\IEEEmembership{Senior member,~IEEE},
Tristan Braud,~\IEEEmembership{Member,~IEEE},
Pan Hui,~\IEEEmembership{Fellow,~IEEE},
Jussi Kangasharju~\IEEEmembership{Member,~IEEE}
\thanks{This work was supported in part by the Academy of Finland in the 5GEAR project, FIT project, and AIDA project, in part by the Hong Kong Research Grants Council in the project 16214817, and FL4IoT project from Huawei, and in part by JSPS KAKENHI under Grants 19H04092, 20H04174, in part by ROIS NII Open Collaborative Research 2020(20FA02).
}
}

\markboth{IEEE Transactions on Intelligent Transportation Systems}
{Shell \MakeLowercase{\textit{Zhou et al.}}: IEEE Transactions on Intelligent Transportation Systems}
\maketitle

\begin{abstract}
The Internet of Vehicles (IoV) enables real-time data exchange among vehicles and roadside units and thus provides a promising solution to alleviate traffic jams in the urban area. Meanwhile, better traffic management via efficient traffic light control can benefit the IoV as well by enabling a better communication environment and decreasing the network load. As such, IoV and efficient traffic light control can formulate a virtuous cycle. Edge computing, an emerging technology to provide low-latency computation capabilities at the edge of the network, can further improve the performance of this cycle. However, while the collected information is valuable, an efficient solution for better utilization and faster feedback has yet to be developed for edge-empowered IoV. To this end, we propose a Decentralized Reinforcement Learning at the Edge for traffic light control in the IoV (DRLE). DRLE exploits the ubiquity of the IoV to accelerate traffic data collection and interpretation towards better traffic light control and congestion alleviation. Operating within the coverage of the edge servers, DRLE aggregates data from neighboring edge servers for city-scale traffic light control. DRLE decomposes the highly complex problem of large area control into a decentralized multi-agent problem. We prove its global optima with concrete mathematical reasoning and demonstrate its superiority over several state-of-the-art algorithms via extensive evaluations.
\end{abstract}


\IEEEpeerreviewmaketitle

\section{Introduction}
\label{sec:intro}

The Internet of Vehicles (IoV)~\cite{contreras2017internet,wu2019integrating,dai2019artificial} allows data exchange among vehicles~(V2V), roadside units~(RSUs)~(V2I), and other commutable devices on roads or remote resources distributed over the Internet. It can facilitate and enable a wide variety of applications such as driving habit monitoring, driving operation recommendation, and emergency notification~\cite{zhang2017safedrive,chou2017car,zhou2018arve}. The IoV leverages an ever-increasing number of vehicles connected to the Internet and has a significant potential to alleviate the continuously rising traffic congestion, which has dramatic consequences on the environment as well as the well-being of citizens. Thanks to its high-speed wireless connectivity, the IoV enables data collection from vehicles in real time~\cite{lu2014connected,wu2020collaborative}.

On a related note, edge computing~\cite{feng2017ave,abbas2017mobile} has emerged as a solution in recent years to extend the capacity of remote cloud services towards nearby end users. Edge computing is at the core of most future networking paradigms as its characteristics make it an ideal candidate for time-sensitive and highly mobile applications such as those encountered in the IoV~\cite{alhilal2020distributed,zhang2018artificial,zhang2019deep,zhang2019edge}. Co-located with base stations or RSUs, edge computing nodes can provide processing of the IoV data and response to traffic jams and anomalies. By connecting traffic signals\footnote{In this work, we use traffic signal and traffic light interchangeably.} to the IoV, it becomes possible to empower signal timing plans with real-time traffic information and exploit the intelligence at the edge to react to unforeseen congestion. However, while the collected information at the edge is valuable, an efficient solution for better utilization and faster feedback has yet to be developed for large-scale edge-empowered IoV. 

Most of the related solutions follow one of the two directions:
\begin{enumerate*}
\item utilize linear programming at intersections for fast adaption of the signal plan~\cite{Barisone2002,dotoli2006signal,coll2013linear},
\item deploy machine learning to directly control the traffic lights or adapt the phase duration~\cite{{balaji2010urban,chu2019multi,li2016traffic}}.
\end{enumerate*}
The first direction lacks exploration or learning abilities and strongly depends on the availability of accurate objective functions and constraints, hence lots of research efforts are put on the second direction, including single-agent~\cite{tsitsiklis1994asynchronous,bingham2001reinforcement} and multi-agent systems~\cite{abdulhai2003reinforcement,de2006reinforcement,chu2019multi,dresner2004multiagent}. Single-agent solutions suffer from huge state space and huge action space. On the other hand, multi-agent solutions can decrease the state and action spaces. Nevertheless, we notice three major concerns regarding the works in this direction as follows.
\begin{enumerate}
\item \textit{Lack of practical solutions} to bridge the technology gap between machine learning algorithms and deployabilities in real-life smart city scenarios.
\item \textit{Lack of solid theoretical analysis} to prove the optimal performances of decentralized training.
\item \textit{Lack of extensive tests with credible simulator platforms} to show the benefit of decentralized learning in traffic light control with reusable results.
\end{enumerate}

In this work, we propose Decentralized Reinforcement Learning at the Edge for traffic light control in the IoV~(DRLE). Following a similar hierarchy with our previous works~\cite{zhou2019enhanced,zhou2019erl}, we build a new system model with a focus on multi-agent training with rich IoV data. This integrated framework leverages the real-time data collection from connected vehicles to optimize traffic light control from the perspective of hierarchical levels. Each level optimizes its coverage with edge servers running a level-specific algorithm, of which one or several key parameters are tuned by the upper level's algorithm in real time. The decentralized architecture of~\sysname relies on a pervasive deployment of edge servers, including signal control units at intersections and edge servers co-located with base stations and aggregation points. 

\sysname decomposes the highly complex problem of large area control into a decentralized multi-agent problem. We prove its global optima with concrete mathematical reasoning~(\cref{sec:algorithm}). We build our algorithm with credible open-sourced platforms~\cite{wu2017flow,moritz2018ray} and reinforcement learning library~\cite{Liang2017RayRA}~(\cref{sec:setup}). We conduct extensive evaluations and demonstrate the superiority of this approach over several state-of-the-art algorithms~(\cref{sec:result}). Specifically,~\sysname decreases convergence time by 65.66\% compared to Proximal Policy Optimization and training steps by 79.44\% compared to Augmented Random Search and Evolutionary Strategies. Besides,~\sysname exponentially reduces the action space and provides comparable traffic control performance within only 1/4 of the training time compared to its centralized counterpart.

The rest of the paper is structured as follows. We give an overview of related works in~\cref{sec:related}. In~\cref{sec:system} we present the system design and traffic model. We describe the theoretical details of our algorithm in~\cref{sec:algorithm} and show our evaluation setup and results in~\cref{sec:setup} and \cref{sec:result}, respectively. Finally,~\cref{sec:conclusion} concludes the paper.

\section{related work}
\label{sec:related}

Researchers have put a lot of effort into optimizing traffic light control. Major solutions include linear programming and machine learning. Meanwhile, recent proposals have started to look at the potential of rich IoV data for traffic management. In this section, we give an overview of related works.

\noindent\textbf{IoV.} Powered by fast-developing vehicular networking techniques, researchers have proposed solutions to utilize rich IoV data for traffic control~\cite{7973459,gerla2014internet}. Kumar et al. proposed to apply ant colony algorithms to help vehicles find the optimal routes~\cite{kumar2018ant}. Darwish et al. focused on real-time big data analytics in the IoV environment powered by fog computing~\cite{darwish2018fog}. Chen et al. targeted enhancing transportation safety and network security by mining effective information from both physical and network data space~\cite{chen2018cognitive}. However, the potential of utilizing rich IoV data specifically for traffic light control has not been fully explored.
\\[2pt]
\noindent\textbf{Liner programming.} Researchers have proposed traffic light control solutions based on traffic models~(microscopic, mesoscopic and macroscopic)~\cite{tyagi2009review,hoogendoorn2001state,eissfeldt2004vehicle}. These solutions utilize linear programming to solve the objective functions~\cite{Barisone2002,dotoli2006signal,coll2013linear}. Although linear programming is straightforward and has low latency, it depends on accurate objective function and constraints. Moreover, it lacks exploration or learning abilities to scale to large-area optimization and complicated scenarios.
\\[2pt]
\noindent\textbf{Reinforcement Learning.} Instead of trying to build explicit traffic flow models, machine learning proposals learn the traffic patterns and achieve optimal policies by iteratively adapting actions with the goal of maximizing the cumulative reward~\cite{abdulhai2003reinforcement,li2016traffic}. Many proposals have applied single-agent or multi-agent reinforcement learning to optimize traffic light control. Single-agent solutions have large state and action spaces thus require large capacity to calculate optimal signal phases~\cite{balaji2010urban}. Therefore, nowadays researchers tend to use multi-agent algorithms to divide larger problems into smaller sub-problems.
For example, Chu et al. proposed in~\cite{chu2016large} to reduce action space by dynamically partitioning the traffic grid into smaller regions and deploying a local agent in each region. They applied multi-agent reinforcement learning to A2C for large-scale traffic light control in~\cite{chu2019multi}. Li et al. used deep Q-learning~(DQL) to control traffic lights and proposed deploying the deep stacked autoencoders (SAE) neural network to reduce the huge state space brought by the tabular Q learning method~\cite{li2016traffic}. Balaji et al. proposed to control traffic lights with distributed agents at each intersection~\cite{balaji2010urban}. El-Tantawy et al. explored coordinated agents to let intersections conduct signal control actions in cooperation with neighbors~\cite{el2013multiagent}. Based on~\cite{chu2016large}, Tan et al. further proposed to concatenate the latent states of local agents to form the global action-value function~\cite{8676356}.

Overall, we find the existing solutions \textbf{fall short} in the following aspects and propose~\sysname to improve those aspects.
\begin{itemize}
\item \textit{Optima proof.} Rarely we find multi-agent approaches for traffic light control mathematically prove the optima of multi-agent reinforcement learning.
\item \textit{The gap between technique and reality.} Most multi-agent learning approaches do not provide deployable solutions to apply the algorithms to real-life smart cities.
\item \textit{Extensive tests with open-source platforms.} Many related works have conducted simulations with vehicular traffic simulators and self-developed machine learning scripts limited to the proposals. We find those results hard to be reused due to lack of credible open-sourced platforms.
\end{itemize}
In this work, we address those shortages by proposing a deployable decentralized solution and prove its optima with mathematical reasoning and extensive reusable tests.
\begin{figure*}[!tb]
	\centering
	\includegraphics[width=\linewidth]{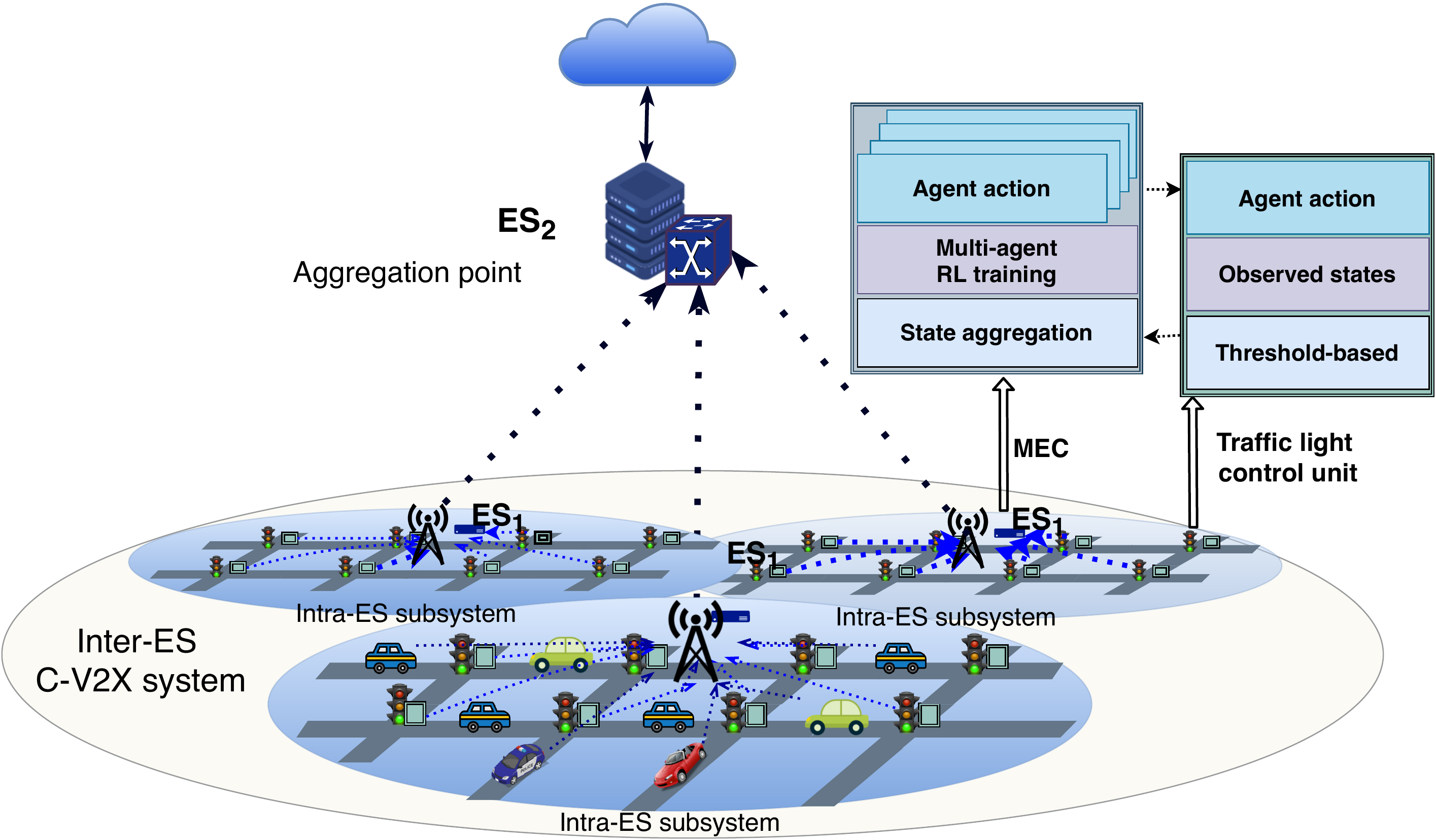}
	\caption{System overview. In each \textit{Intra-ES} subsystem, an $\text{ES}_1$~(MEC) uses decentralized multi-agent learning to train the collected IoV data and sends commands to control the traffic lights. $\text{ES}_2$ tunes the parameters of the multi-agent algorithms running in each $\text{ES}_1$~(\textit{Intra-ES} subsystem).}
	\label{fig:system_overview}
\end{figure*}
\begin{figure}[!tb]
	\centering
	\includegraphics[width=.9\linewidth]{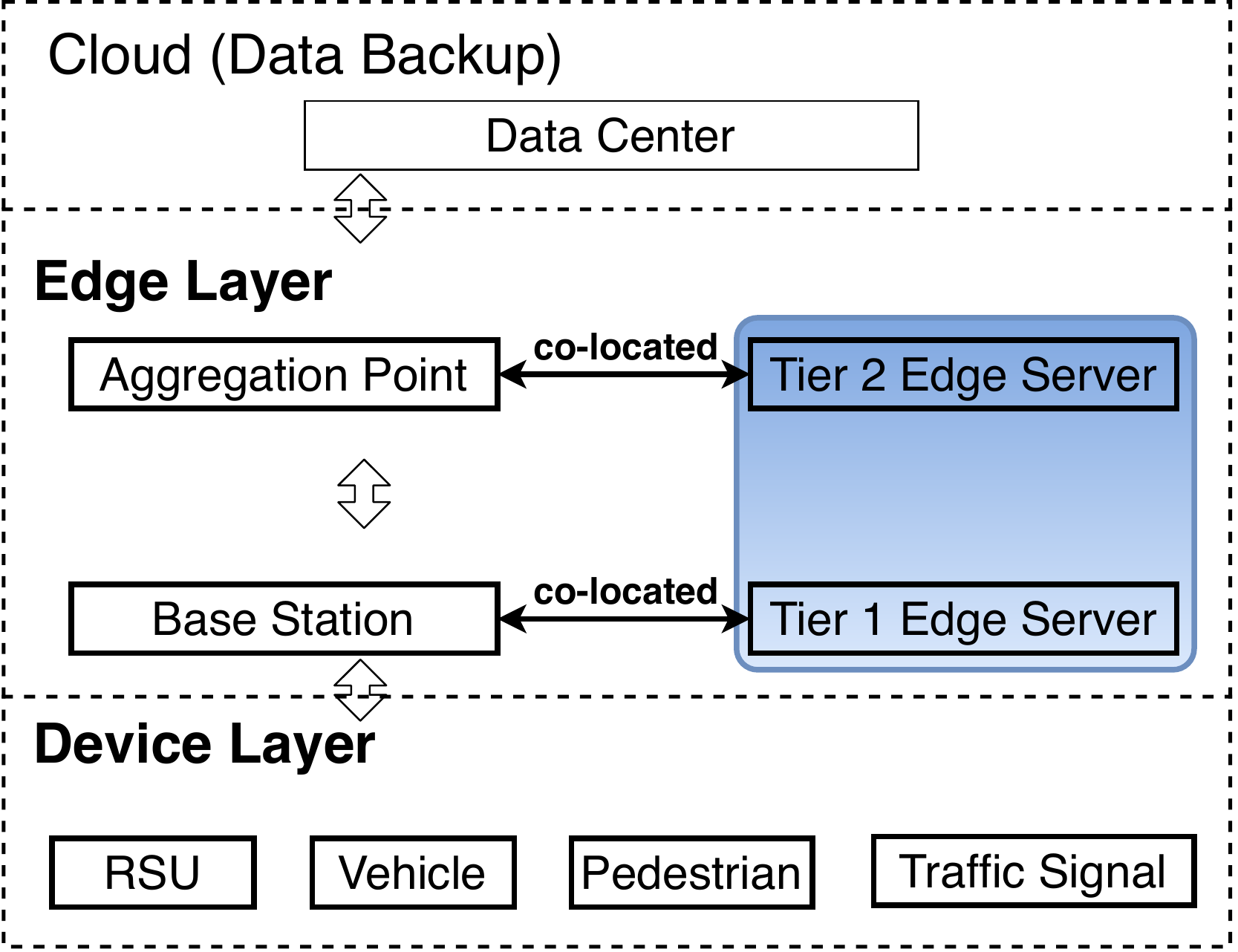}
	\caption{System layers.}
	\label{fig:system_arc}
\end{figure}

\section{System}
\label{sec:system}
In this section, we first present the design of \sysname and the communication mechanism. Then, we describe the traffic model as the basis of the algorithm.
\subsection{System Design}
\label{subsec:architecture}
\sysname revolves around the device layer and the edge layer, as shown in Fig.~\ref{fig:system_overview} in detail and Fig.~\ref{fig:system_arc} from high level. The device layer includes the vehicles, traffic signals, pedestrian devices, RSUs and other devices involved in the IoV. In the rest of this paper, we assume each traffic signal is connected with a \textit{control unit}. Each control unit employs video cameras facing all directions to collect real-time traffic data and transmits it via wireless network to a nearby edge server. The edge layer hosts the edge servers in two tiers. The first tier servers~($\text{ES}_1$) are co-located with the base stations at the radio access network. The second tier servers~($\text{ES}_2$) are co-located with the aggregation points in the core network. This scheme agrees with the Multi-access Edge Computing standard proposed by the European Telecommunications Standards Institute. Please refer to~\cite{zhou2019enhanced} for the detailed edge server deployment strategy. Each $\text{ES}_2$ collects data from nearby $\text{ES}_1$ and provides a larger scale of service and sends backup to the cloud data center. Next, we briefly describe the communication mechanism and overhead.

\subsection{Communications}
\label{subsec:comm_analysis}

Each $\text{ES}_1$ feeds the collected data into local reinforcement learning and sends the actions to the signal control units in real time. An $\text{ES}_1$ collects the data by communicating with the devices (connected vehicles, signal control units and RSUs) via Cellular vehicle-to-Everything (C-V2X)~\cite{papathanassiou2017cellular} and operates within a coverage defined by the range of its co-located base station. For simplicity, we do not consider RSUs and pedestrians in the rest of the paper.

Actions are primarily changing lights commands which can be encapsulated in small data packets. These packets are sent only to a limited number of signals, hence the majority of data transmission is between the connected vehicles and $\text{ES}_1$ via Vehicle-to-Network~(V2N). As suggested by the 3GPP standard~\cite{3gppv2n}, each message should be sent at a frequency between 0.1\,Hz and 1\,Hz with a payload between 50 bytes and 300 bytes. We assume a message frequency of 1\,Hz to align with the agent learning rate in the evaluation~(see~\cref{sec:setup}). We assume that each message has a size of maximum transmission unit (1500 bytes), which is sufficient to contain the required data~(speed, location, and direction of travel of the vehicle). Note that other transmissions mechanisms such as V2V and V2I are performed via a different radio (e.g. ITS 5.9\,GHz) than V2N~(licensed mobile bands, e.g. 700\,MHz) and are not necessary for~\sysname.

Since the transmissions between the vehicles and the $\text{ES}_1$ base stations are only one-hop and unidirectional~(uplink), the communications require a small networking capacity. Hence, it has a limited overhead and impact on the overall C-V2X environment. We present a preliminary evaluation in~\cref{subsec:comm} and show that the end-to-end delay consisting of vehicle-to-$\text{ES}_1$ transmission delay and $\text{ES}_1$-to-signal transmission delay is much smaller than a training step duration~(1 second).\footnote{A training step is the period over which one gradient update happens. We define each step as a 1-second period. However, the delay of each gradient update is at the millisecond level. In other words, each gradient update takes several milliseconds then waits until the next second to conduct the next update.}

\subsection{Traffic Model and Problem}
\label{subsec:vmodel}
We assume the traffic lights in the urban area pertain to a common signal timing plan characterized by a fixed cycle containing a fixed number of phases. A phase refers to the time duration of the green lights for a given direction. Let $\mathcal{L}$ be the set of links in a signalized urban area. Then, $\mathcal{L}_{(\mathrm{in})}$~($\mathcal{L}_{(\mathrm{out})}$) is the set of input~(output) links of the area. 
To allow smooth driving experience, we introduce two parameters, i.e, the number of halting vehicles and speed-lag.
A vehicle moves at speeds slower than 0.1\,m/s is considered halting.
Speed-lag is defined as the speed difference between the actual driving speed and the maximum speed permitted by statute. We use \emph{speed-lag} instead of the commonly used \emph{speed} to address the maximum speed limit in reality.
We formulate the problem as to minimize the overall number of halting vehicles and speed-lag in the area over the whole optimization horizon. As for the coverage of an $\text{ES}_1$, the objective of the optimization problem can be formulated as follows,
\begin{align}
\label{eq:internal}
    \bar{P} = \lim_{K\to\infty} \textsf{E}\!\Bigg[\frac{1}{K} \sum_{k=1}^{K}
    \sum_{ m \in L_{(\mathrm{in})} \setminus L_{(\mathrm{out})}} \left( -w_1 \cdot H_m^k - w_2 \cdot \triangle{V}_m^k\right) \Bigg],
\end{align}
which can also be approximated as
\begin{align}
\label{eq:internal2}
   P =    
   \textsf{E}\!\Bigg[(1 - \gamma) \cdot \sum_{k=1}^{\infty} (\gamma)^{k - 1} \cdot 
    \sum_{\mathclap{m \in \mathcal{L}_{(\mathrm{in})} \setminus \mathcal{L}_{(\mathrm{out})}} }
    \left(-w_1 \cdot H_m^k - w_2 \cdot \triangle{V}_m^k\right) \Bigg],              
\end{align}
\\
if the discount factor $\gamma \in [0, 1)$ approaches 1. Herein, $H_m^k$ is the number of halting vehicles and $\triangle{V}_m^k$ is the average speed-lag of the vehicles in lane $m$ at the beginning of the $k$th cycle. $w_1$ and $w_2$ are two positive weighting constants. Although the model describes the problem from a global view, each internal intersection can have a varied impact on the optimization. Therefore, it is better to disassemble the optimization objective of multiple intersections, namely,
\begin{align}
\label{eq:intersection}
  P = \textsf{E}\!\Bigg [ (1 - \gamma) \cdot \sum_{k=1}^{\infty} (\gamma)^{k - 1} \cdot
    &\sum_{c=1}^{C} \sum_{m \in \mathcal{L}^c_{(\mathrm{in})} \setminus \mathcal{L}^c_{(\mathrm{out})}}\\ &\left(-w_1 \cdot H_m^k
    - w_2 \cdot \triangle{V}_m^k\right) \Bigg ].\nonumber
\end{align}
where $c$ is an intersection index, $C$ refers to the total number of intersections in the area, $\mathcal{L}^c_{(\mathrm{in})}$ ($\mathcal{L}^c_{(\mathrm{out})}$) represents the set of input (output) links of intersection $c$.
In this work, we propose to decompose the optimization problem and solve it with a decentralized reinforcement learning algorithm as described in \cref{sec:algorithm}.

\section{Algorithm}
\label{sec:algorithm}
\subsection{Hierarchical Algorithm}
The system includes three parallel and interactive algorithms running in three levels, i.e., \textit{Intersection}, \textit{Intra-ES} and \textit{Inter-ES}, performed by signal control units, $\text{ES}_1$ and $\text{ES}_2$, respectively~(Fig.~\ref{fig:system_overview}). At \textit{Intersection} level, each signal adapts its phases with a threshold-based algorithm. The threshold can be the average queuing length or the average space headaway. At \textit{Intra-ES} level, each $\text{ES}_1$ runs a multi-agent reinforcement learning algorithm to participate in the traffic light switching directly. At \textit{Inter-ES} level, an $\text{ES}_2$ runs a threshold-based algorithm to optimize the urban traffic by tuning the reinforcement learning rate in each $\text{ES}_1$. We briefly describe \textit{Intersection} level and \textit{Intra-ES} level and detail the decomposed reinforcement learning algorithm at \textit{Intra-ES} level.
\\[2pt]
\noindent\textbf{\textit{Intersection} level.} The control unit of each intersection signal runs a threshold-based algorithm similar with~\cite{Krajzewicz2005}. 
Each control unit employs cameras to capture the traffic jam and adapts the phase duration based on predefined parameters. When the parameters surpass the thresholds, the control unit extends the phase of the green lights for the more jammed direction~(e.g., with longer queues or smaller space headway), and decreases a equivalent length of the green phases for the other direction.
\\[2pt]
\noindent\textbf{\textit{Inter-ES} level.} An $\text{ES}_2$ tunes the urban traffic by adapting the reinforcement learning rate in each $\text{ES}_1$ based on a threshold-based algorithm similar to~\cite{Krajzewicz2005}.
\\[2pt]
\noindent\textbf{\textit{Intra-ES} level.} Each $\text{ES}_1$ optimizes the internal traffic within its coverage defined by the co-located base station by switching the traffic lights directly. We adopt DQL to provide an adaptive algorithm to respond to the dynamically changing traffic condition. The advantage of Q-learning for traffic light control is described in more detail in a study by Abdulhai et al.~\cite{abdulhai2003reinforcement}, including not requiring a prespecified model of the environment and being adaptive and unsupervised. We define the state $\boldsymbol{s}^k$, action $\boldsymbol{a}^k$ and reward $R^k$ as follows.
\begin{itemize}
    \item \noindent$\boldsymbol{s}^k$ is the state at the $k$th cycle, including the number of halting vehicles $\boldsymbol{H}^k = \{H_m^k| m = 1, \cdots, M\}$, the speed-lag of the vehicles $\boldsymbol{\triangle{V}}^k = \{\triangle{V}_m^k| m = 1, \cdots, M\}$, and the traffic light states $\boldsymbol{\theta}^k = \{\boldsymbol{\theta}_c^k| c = 1, \cdots, C\}$, where $M$ is the total number of the lanes in the area.
    \item \noindent$\boldsymbol{a}^k$ is the action operated by the agents after observing $\boldsymbol{s}^k$. More specifically, $\boldsymbol{a}^k =\{a_c^k| c = 1, \cdots, C\}$, where $a_c^k \in \{0, 1\}$ represents the decision of switching the traffic light.
    \item $\boldsymbol{R^k}$ is the reward defined as the additive inverse of the average speed-lag and number of halting vehicles, namely, $R^k = \sum_{m = 1}^M (-w_1 \cdot H_m^k - w_2 \cdot \triangle{V}_m^k)$.
\end{itemize}

\textbf{Agent Goal.} The overall goal of \sysname is to optimize the light control to smooth the traffic. To do so, the agents of an $\text{ES}_1$ need to find a control policy $\pi$ that maximizes
\begin{equation}
\label{eq:reward1}
Q^{\pi}(\boldsymbol{s},\boldsymbol{a})= (1 - \gamma) \cdot \textsf{E}\! \left[ \sum_{k=1}^{\infty} (\gamma)^{k - 1} \cdot R^k | \boldsymbol{s}^1 =\boldsymbol{s}, \boldsymbol{a}^1 = \boldsymbol{a}\right],
\end{equation}
which is also termed a Q-function. A control policy $\pi$ can be defined as a mapping by $\boldsymbol{a} = \pi(\boldsymbol{s})$. To put it another way, the goal is to solve
\begin{equation}
\label{eq:reward2}
\pi^*=\underset{\pi}{\argmax}~Q^{\pi}(\boldsymbol{s},\pi(\boldsymbol{s})), \forall \boldsymbol{s}.
\end{equation}
For notational convenience, we denote $Q(\boldsymbol{s}, \boldsymbol{a}) = Q^{\pi^*}(\boldsymbol{s}, \boldsymbol{a})$, $\forall (\boldsymbol{s}, \boldsymbol{a})$. Using the state-action-reward-state-action (SARSA) algorithm~\cite{sutton1998reinforcement}, the optimal Q-function can be found in an iterative on-policy manner. The centralized decision making at a cycle is executed at the signals independently, based on which we linearly decompose the Q-function,
\begin{align}\label{QFuncDeco}
   Q(\boldsymbol{s}, \boldsymbol{a}) = \sum_{c = 1}^C Q_c(\boldsymbol{s}, \boldsymbol{a}_c),
\end{align}\\
where $Q_c(\boldsymbol{s}, \boldsymbol{a}_c)$ is defined to be the per-signal Q-function given by
\begin{equation}
\label{eq:reward3}
Q_c(\boldsymbol{s}, \boldsymbol{a}_c) = (1 - \gamma) \cdot \textsf{E}\! \left[ \sum_{k=1}^{\infty} (\gamma)^{k - 1} \cdot R_c^k | \boldsymbol{s}^1 =\boldsymbol{s}, \boldsymbol{a}_c^1 = \boldsymbol{a}_c\right].
\end{equation}
Herein, $\boldsymbol{a}_c^k$ and $R_c^k$ are, respectively, the joint action and the reward for the agent at the intersection $c$ at the cycle $k$. We emphasize that the decision makings across the cycles are performed at each signal in accordance with the optimal control policy implemented by the $\text{ES}_1$. In other words, $\forall \boldsymbol{s}$,
\begin{align}\label{optiActi}
  \pi^*(\boldsymbol{s}) = \underset{\boldsymbol{a} = (\boldsymbol{a}_c: c = 1, \cdots, C)}{\arg\max} \sum_{c = 1}^C Q_c(\boldsymbol{s}, \boldsymbol{a}_c).
\end{align}
\subsection{Optimization and Convergence Guarantee}
\noindent\textbf{Theorem 1} (Optimization Guarantee).
\emph{The linear Q-function decomposition approach as in Eq.~(\ref{QFuncDeco}) asserts the expected optimal long-term performance.}
\\[2pt]
\textbf{Proof.}
For the Q-function of a centralized decision making $\boldsymbol{a}$ under a global network state $\boldsymbol{s}$, we have
\begin{align}\label{QDecoOpti}
     Q(\boldsymbol{s}, \boldsymbol{a})
 & = (1 - \gamma) \cdot \textsf{E}\!\left[\sum_{k = 1}^\infty (\gamma)^{k - 1} \cdot
     R^k | \boldsymbol{s}^1 = \boldsymbol{s}, \boldsymbol{a}^1 = \boldsymbol{a}\right]                                          \nonumber\\
 & = (1 - \gamma) \cdot \textsf{E}\!\left[\sum_{k = 1}^\infty (\gamma)^{k - 1} \cdot
     \sum_{c = 1}^C R_c^k | \boldsymbol{s}^1 = \boldsymbol{s}, \boldsymbol{a}^1 = \boldsymbol{a}\right]                         \nonumber\\
 & = \sum_{c = 1}^C (1 - \gamma) \cdot \textsf{E}\!\left[\sum_{k = 1}^\infty (\gamma)^{k - 1} \cdot
     R_c^k | \boldsymbol{s}^1 = \boldsymbol{s}, \boldsymbol{a}^1 = \boldsymbol{a}\right]                                        \nonumber\\
 & = \sum_{c = 1}^C Q_c(\boldsymbol{s}, \boldsymbol{a}_c),
\end{align}
which completes the proof.
\hfill$\Box$

Therefore, instead of learning the Q-function, the SARSA updating rule is slightly adapted for each lane to
\begin{align}\label{perVUEQLearRule}
     Q_c^{k + 1}(\boldsymbol{s}, \boldsymbol{a}_c)
 & = \left(1 - \alpha^k\right) \cdot Q_c^k\!\left(\boldsymbol{s}, \boldsymbol{a}_c\right)           \nonumber\\
 & + \alpha^k \cdot \left((1 - \gamma) \cdot R_c^k +
     \gamma \cdot Q_c^k\!\left(\boldsymbol{s}', \boldsymbol{a}_c'\right)\right),
\end{align}
where $\alpha^k \in [0, 1)$ is the learning rate. Theorem 2 ensures the convergence of the decentralized learning process.

\noindent\textbf{Theorem 2} (Convergence Guarantee).
\emph{The sequence $\{(Q_c^k(\boldsymbol{s}, \boldsymbol{a}_c): \forall (\boldsymbol{s}, \boldsymbol{a}_c), \forall c \in \{1, \cdots, C\}): k\}$ by Eq.~(\ref{perVUEQLearRule}) surely converges to the per-signal Q-functions $(Q_c(\boldsymbol{s}, \boldsymbol{a}_c): \forall (\boldsymbol{s}, \boldsymbol{a}_c), \forall c \in \{1, \cdots, C\})$, if and only if for each signal $c \in \{1, \cdots, C\}$, the $(\boldsymbol{s}, \boldsymbol{a}_c)$-pairs are visited for an infinite number of times.}

\noindent\textbf{Proof.}
Since the per-signal Q-functions are learned simultaneously, we consider monolithic updates during the decentralized learning process. That is, the iterative rule in Eq.~\eqref{perVUEQLearRule} can then be encapsulated as
\begin{align}\label{SARSA}
 & \sum_{c= 1}^C Q_c^{k + 1}(\boldsymbol{s}, \boldsymbol{a}_c) =
   \left(1 - \alpha^k\right) \cdot \sum_{c = 1}^C Q_c^k\!\left(\boldsymbol{s}, \boldsymbol{a}_c\right) + \nonumber\\
 & \alpha^k \cdot \left((1 - \gamma) \cdot \sum_{c= 1}^C R_c +
   \gamma \cdot \sum_{c = 1}^C Q_c^k(\boldsymbol{s}', \boldsymbol{a}_c')\right).
\end{align}\\
From both sides of Eq.~\eqref{SARSA}, subtracting the sum of per-signal Q-functions leads to
\begin{align}\label{SARSAv2}
 & \sum_{c = 1}^C Q_c^{k + 1}(\boldsymbol{s}, \boldsymbol{a}_c) - \sum_{c = 1}^C Q_c(\boldsymbol{s}, \boldsymbol{a}_c) =    \nonumber\\
 & \left(1 - \alpha^k\right) \cdot \left(\sum_{c = 1}^C Q_c^k(\boldsymbol{s}, \boldsymbol{a}_c) -
                                         \sum_{c = 1}^C Q_c(\boldsymbol{s}, \boldsymbol{a}_c)\right) +                    \nonumber\\
 & \alpha^k \cdot T^k(\boldsymbol{s}, \boldsymbol{a}_c),
\end{align}
where
\begin{align}
      T^k(\boldsymbol{s}, \boldsymbol{a}_c)
 & = (1 - \gamma) \cdot \sum_{c = 1}^C R_c                                                          \\
 & + \gamma \cdot \max_{\boldsymbol{a}''} \sum_{c = 1}^C Q_c^k\!\left(\boldsymbol{s}', \boldsymbol{a}_c''\right) -
     \sum_{c = 1}^C Q_c(\boldsymbol{s}, \boldsymbol{a}_c))                                                                            \nonumber\\
 & + \gamma \cdot \left(\sum_{c = 1}^C Q_c^k(\boldsymbol{s}', \boldsymbol{a}_c') -
     \max_{\boldsymbol{a}''} \sum_{c = 1}^C Q_c^k(\boldsymbol{s}', \boldsymbol{a}_c'')\right).                              \nonumber
\end{align}\\
We let $\Delta^k$ denote the history for the first $k$ cycles during the decentralized learning process. The per-signal Q-functions are $\Delta^k$-measurable, thus both $(\sum_{c = 1}^C Q_c^{k + 1}(\boldsymbol{s}, \boldsymbol{a}_c) - \sum_{c = 1}^C Q_c(\boldsymbol{s}, \boldsymbol{a}_c))$ and $T^k(\boldsymbol{s}, \boldsymbol{a}_c)$ are $\Delta^k$-measurable. We then attain Eq.~\eqref{sarsaconv}, where $\|\cdot\|_\infty$ is the maximum norm of a vector and (a) is due to the convergence property of the standard Q-learning. We are now left with verifying that $\|\textsf{E}[\gamma \cdot (\sum_{c = 1}^C Q_c^k(\boldsymbol{s}', \boldsymbol{a}_c') - \max_{\boldsymbol{a}''}$ $\sum_{c = 1}^C Q_c^k(\boldsymbol{s}', \boldsymbol{a}_c'')) | \Delta^k]\|_\infty$ converges to zero, which establishes the following: i) an $\epsilon$-greedy policy is deployed for the exploration-exploitation trade-off during decision-making; ii) the per-signal Q-function values are upper bounded; and iii) both the global network state and the decision-making spaces are finite. Thus the convergence of the decentralized learning is ensured.
\hfill$\Box$
\begin{figure*}[b]
\begin{align}\label{sarsaconv}
      & \left\|\textsf{E}\!\!\left[T^k(\boldsymbol{s}, \boldsymbol{a}_c) | \Delta^k\right]\right\|_\infty                 \\
 \leq & \left\|\textsf{E}\!\!\left[(1 - \gamma) \cdot \sum_{c = 1}^C R_c +
               \gamma \cdot \max_{\boldsymbol{a}''} \sum_{c = 1}^C Q_c^k(\boldsymbol{s}', \boldsymbol{a}_c'') -
               \sum_{c = 1}^C Q_c(\boldsymbol{s}, \boldsymbol{a}_c) | \Delta^k\right]\right\|_\infty +                        \nonumber\\
      & \left\|\textsf{E}\!\!\left[\gamma \cdot \left(\sum_{c = 1}^C Q_c^k(\boldsymbol{s}', \boldsymbol{a}_c') -
               \max_{\boldsymbol{a}''} \sum_{c= 1}^C Q_c^k(\boldsymbol{s}', \boldsymbol{a}_c'')\right) |
               \Delta^k\right]\right\|_\infty                                                                               \nonumber\\
 \overset{\mbox{(a)}}{\leq}
      & \gamma \cdot \left\|\sum_{c = 1}^C Q_c^k(\boldsymbol{s}, \boldsymbol{a}_c) -
                            \sum_{c = 1}^C Q_c(\boldsymbol{s}, \boldsymbol{a}_c)\right\|_\infty +
        \left\|\textsf{E}\!\!\left[\gamma \cdot \left(\sum_{c = 1}^C Q_c^k(\boldsymbol{s}', \boldsymbol{a}_c') -
               \max_{\boldsymbol{a}''} \sum_{c = 1}^C Q_c^k(\boldsymbol{s}', \boldsymbol{a}_c'')\right) |
               \Delta^k\right]\right\|_\infty                                                                               \nonumber
\end{align}
\hrule
\end{figure*}

\textbf{Takeaway.} The core contribution of the algorithm, i.e. decentralization, lies in the action selection process, during which the algorithm selects the optimal joint action that maximizes the sum of Q-values of all agents, thus ensuring global optima (Theorem 1 and Theorem 2). The major advantage in comparison with traditional centralized reinforcement learning, is that the number of joint actions grows linearly instead of exponentially as the number of involved agents (i.e., the traffic signals) increases. It is noteworthy that the training is done in a centralized fashion to allow efficient cooperative training across the multiple agents.

\section{Experiment Setup}
\label{sec:setup}
We conduct the training and the tests on a MSI GS65 Stealth 8SG equipped with a 6-core I7-8750H CPU, 32GB of memory, and an Nvidia RTX 2080 Max-Q GPU. We build the learning algorithms (open sourced at~\cite{code}) based on \texttt{RLlib}~\cite{Liang2017RayRA}, an open-sourced library for reinforcement learning that can easily be scaled by increasing the number of workers~\cite{mnih2013playing}. We use its underlayer \texttt{Ray} framework~\cite{moritz2018ray} to accelerate the decentralized algorithm training.

Besides Deep Q-Network~(\emph{DQN}), we also build and test distributional DQN~(\emph{dDQN}) proposed by Bellemare et al. in 2017~\cite{10.5555/3305381.3305428}, which learns a categorical distribution of discounted returns instead of estimating the mean. To provide better performance, we leverage the values of several parameters explored by the well-acknowledged work, \emph{Rainbow}~\cite{hessel2018rainbow}, including learning rate, epsilon, and softmax cross entropy for \emph{dDQN}. We define and evaluate the algorithms and policies utilizing \texttt{Flow}~\cite{wu2017flow}, a Python library that provides the interface between \texttt{RLlib} and \texttt{SUMO}~\cite{SUMO2012}, a microscopic simulator for traffic and vehicle dynamics. Next we describe the detailed settings.
\subsection{Simulation Setup}
\label{subsec:setup}
To extensively test our proposal, we evaluate sets of tests with different algorithms and policies in different scenarios.
\\[2pt]
\noindent\textbf{Map.} Most related works on traffic light control do experiments on grid-like maps such as the urban Manhattan grid scenario used by 3GPP~\cite{3gppv2x}. We also follow this setup and deploy the tests on $n \times n$ grid road maps that consist of $n \times n$ typical four-way, traffic-light-controlled intersections. Each intersection allows vehicles to flow either horizontally or vertically. If the light is green, it transitions to yellow for two seconds before switching to red for the purpose of safety. Each lane has one signal only, and the signal phases of an intersection in clockwise order consist of \textit{GrGr}, \textit{yryr}, \textit{rGrG}, and \textit{ryry}, of which \textit{G, y, r} refer to green, yellow and red, respectively. The traffic lights can be switched only after $3$ seconds to prevent flickering.
\\[2.5pt]
\noindent\textbf{Scale.} In this work, each $\text{ES}_1$ collects data and control the traffic lights within the coverage defined by its co-located base station. To be more realistic, we investigate the coverage range of LTE cell towers via a crowdsourced cellular tower and coverage mapping service\footnote{https://www.cellmapper.net/}. It shows that the coverage range of an LTE cell tower on B7~(2600MHz) is typically around 5 to 10 blocks and 10 to 20 intersections in a European city center area~(Helsinki, Finland). Therefore, we focus the simulation at the scale of of $5\times5$ intersections to cover common scenarios. For the completeness of the work, we also test larger scales such as $10\times10$ and $15 \times15$.\\[2.5pt]
\begin{figure}
		\centering
		\includegraphics[width=.9\linewidth]{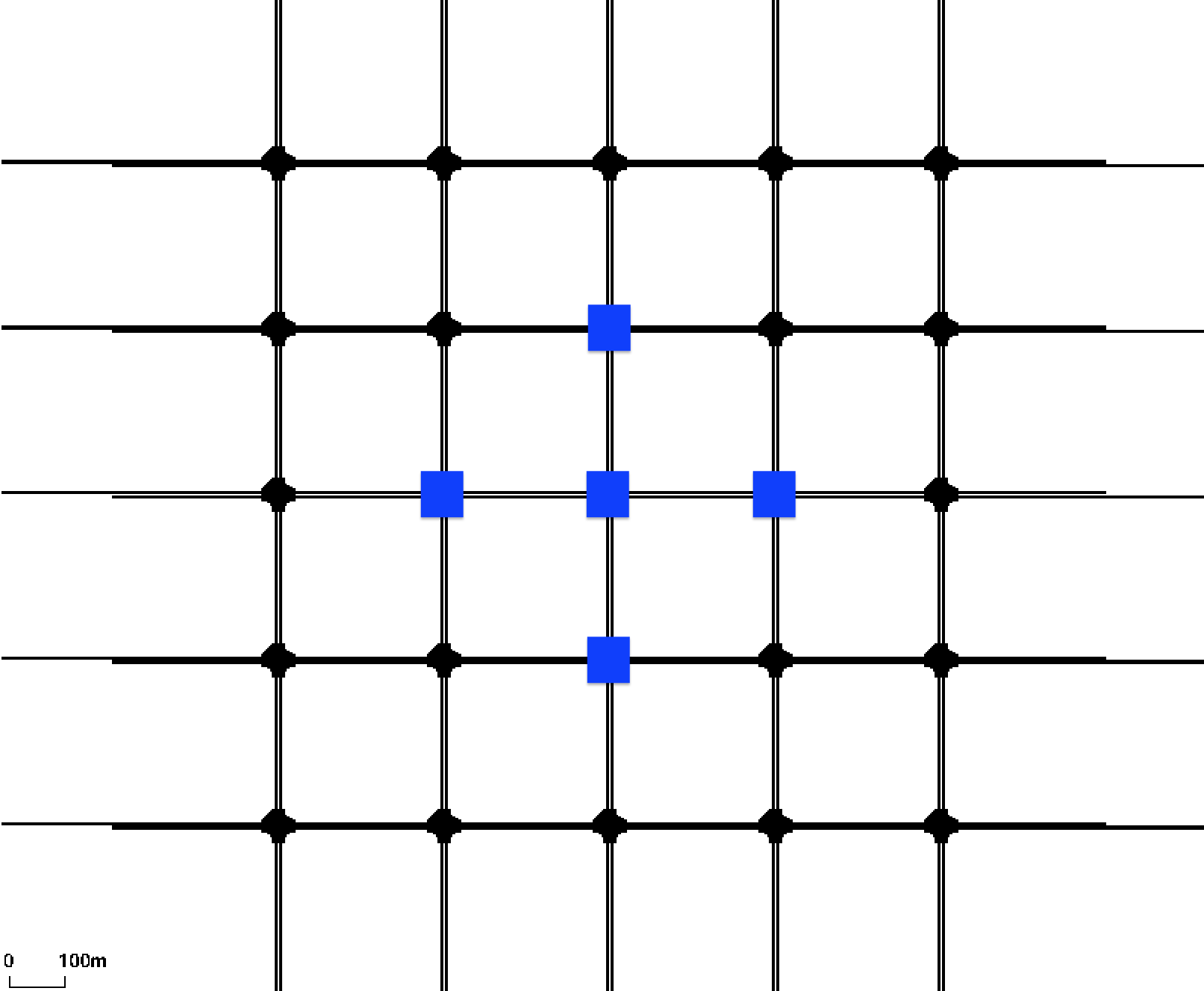}
		\caption{Central (blue squares) and edge (black circles) intersections.}
		\label{fig:map}
\end{figure}
\noindent\textbf{Traffic.} Vehicles enter the map from all outer edges at a predefined rate, i.e., 360 \textit{vehicle/hour/edge}. At such a rate, 7200 vehicles enter a $5\times5$ grid map from the 20 outer edges in an hour. To simplify the problem, vehicles travel straight on their paths. Each vehicle is driven following a basic \texttt{SUMO}-built-in car following model, of which the minimum gap between successive vehicles, maximum speed limit and deceleration ability are set to 2.5\,m, 60\,m/s and 7.5\,m/s$^2$, respectively.

\subsection{Training Configurations}
\label{subsec:train}
We test the learning algorithms utilizing single- and multi-agent training. We use multi-agent \emph{DQN} and \emph{dDQN} as the targeting algorithms to prove our mathematical reasoning in~\cref{sec:algorithm}. Considering intersections with different centralities may have influences on the traffic performance, we define two kinds of policies with corresponding reward definitions as follows.
\\[2.5pt]
\noindent\textbf{Policy.} We test the system with two policies, i.e., \emph{SharedPolicy} and \emph{MultiPolicy}, to address the different centralities of the intersections. We call the signals with higher centralities ``central nodes", as indicated by the blue squares residing in the central area of Fig.~\ref{fig:map}. We call the others ``edge nodes", as indicated by the black circles. \emph{SharedPolicy} lets the agents at all intersections share the same policy. \emph{MultiPolicy} lets the agents at ``central nodes" share a \emph{central} policy while the other agents share a \emph{edge} policy.

\noindent\textbf{Reward.} We define the reward based on the average \emph{speed-lag} and \emph{the number of halting vehicles}~(\cref{sec:system}). Specifically, the reward of \emph{SharedPolicy} is defined as
\begin{align}\label{eq:share-reward}
   R= -w_1 \cdot H - w_2 \cdot \triangle{V}
\end{align}
where $R$, $H$, $\triangle{V}$ indicate the reward, number of halting vehicles, and average speed-lag, respectively. The rewards of \emph{MultiPolicy} for the agents at ``central nodes" and ``edge nodes" are defined as
\begin{align}\label{eq:central-reward}
   R_{central}= -w^c_1 \cdot H - w^c_2 \cdot \triangle{V}
\end{align}
\begin{align}\label{eq:edge-reward}
   R_{edge}=  -w^e_1 \cdot H - w^e_2 \cdot \triangle{V}
\end{align}
where $w^c_1$, $w^c_2$, $w^e_1$, $w^e_2$ denote the weights to differentiate the penalties that central and edge nodes receive. In this paper, we set $w^c_1 > w^e_1, w^c_2>w^e_2$, to give central nodes higher penalties for poor performance. 

\subsection{Benchmarks}
\label{subsec:benchmark}
We compare the performance of \emph{DQN} and \emph{dDQN} with multiple algorithms as follows.
\begin{itemize}
\item \textit{Static} simply lets traffic lights deploy pre-defined static phases.
\item \textit{Actuated} is a common light control scheme in Germany. It works by either prolonging traffic phases upon detecting a continuous traffic stream, or switching to the next phase upon detecting a sufficient time gap between successive vehicles~\cite{sumolight}.
\item \textit{Augmented Random Search (ARS)} is an improved version of Basic Random Search~(BRS), proposed by Mania et al. in 2018~\cite{ars}.
\item \textit{Evolutionary Strategies (ES)} is one of the OpenAI solutions proposed by Salimans et al. in 2017~\cite{salimans2017evolution}.
\item \textit{Proximal Policy Optimization (PPO)} is a popular gradient-based policy optimization algorithm proposed by Schulman et al. in 2017~\cite{schulman2017proximal}. It uses multiple epochs of mini-batch updates, and an MLP with tanh non-linearity to compute a value function baseline.
\end{itemize}
We also test \emph{dDQN} using all parameter values deployed in \emph{Rainbow}~\cite{hessel2018rainbow} to see whether it applies to our use case. Table~\ref{tab:parameter} shows the hyper-parameters of the benchmark algorithms, \emph{DQN} and \emph{dDQN}.
\renewcommand{\arraystretch}{1.4}
\begin{table}[t]
 \centering
 \caption{Hyper-parameters.}
 \label{tab:parameter}
 \begin{tabular}{c|l}
    \specialrule{1.3pt}{1pt}{1pt}
    Algorithm & Selected Hyper-parameters  \\ \hline
    Augmented Random Search  & SGD step size = 0.2\\
    &Noise standard deviation =  0.2 \\\hline
    Evolutionary Strategies & Adam step size = 0.02\\
    &Noise standard deviation = 	0.02 \\\hline
    Proximal Policy Optimization & $\lambda(\text{GAE})=0.3$\\
    &Clipping $\epsilon =0.3$ \\
    & Adam step size $=\num{5e-5}$\\
    &Minibatch size $=128$ \\
    &Hiddens [100, 50, 25] \\
    &$\gamma(\text{Discount})=0.999$\\ \hline
    Deep Q-Network &Adam $\epsilon = \num{1.5e-4}$ \\
    &Learning rate $\alpha=\num{6.25e-05}$\\
    &Hiddens [256,256] \\
    &Dimension $=84$\\ 
     &Train batch size $= 1000$ \\ 
    \hline
    Distributed DQN & Number of atoms $=51$\\
    &Min/max values: $[-10,10]$\\
    &Target network update freq. $=8000$\\
      \specialrule{1.3pt}{1pt}{1pt}
  \end{tabular}
\end{table}
\section{Test Results}
\label{sec:result}

\subsection{Training Results}
\label{subsec:training}
\begin{figure*}[!tb]
\centering
\captionsetup{justification=centering}
	\begin{subfigure}[t]{0.31\textwidth}
		\includegraphics[width=\textwidth]{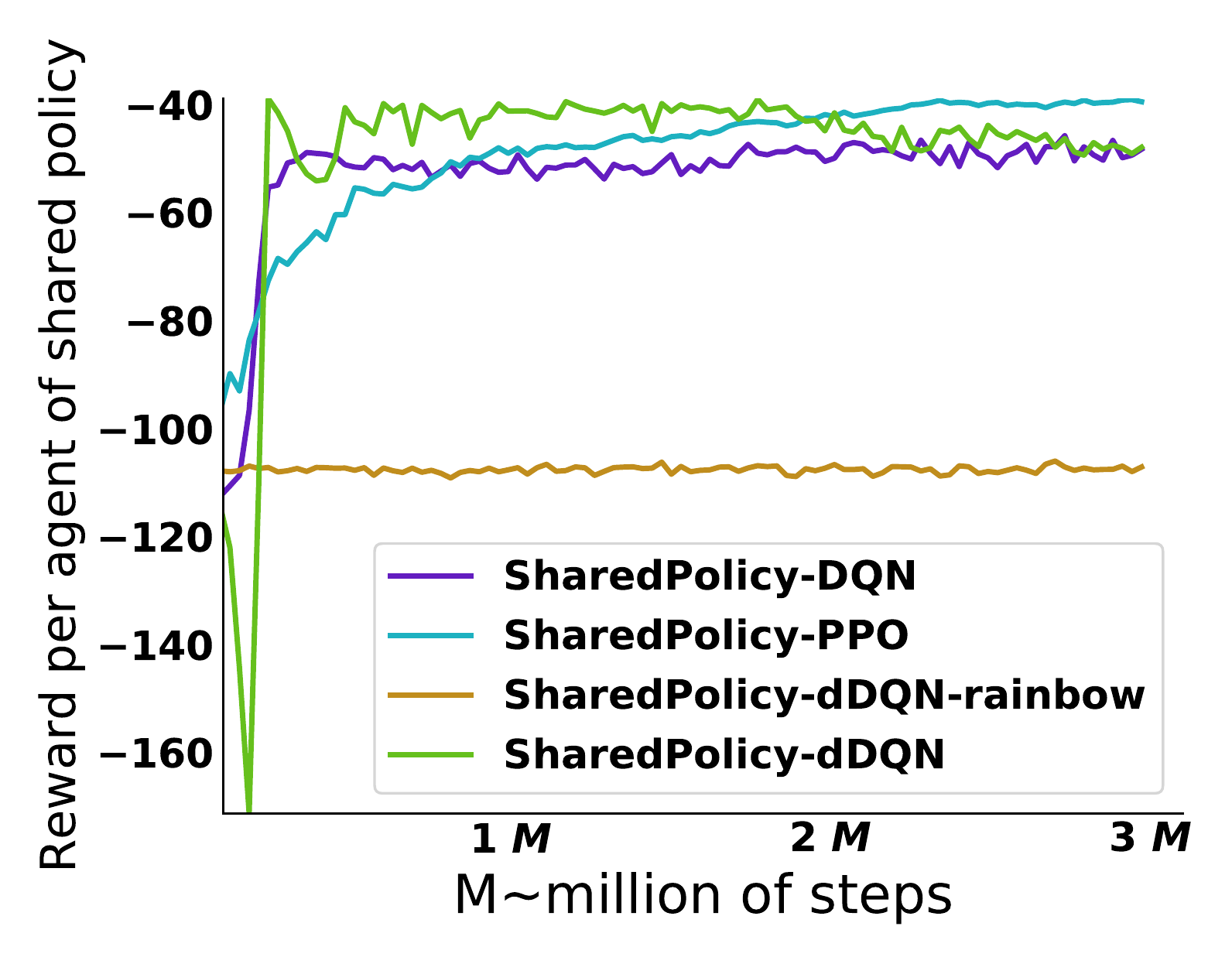}
		\caption{Average reward per agent for \emph{SharedPolicy}~(reward defined in Eq.~(\ref{eq:share-reward})).}
		\label{fig:policy-shared}
	\end{subfigure}
	\begin{subfigure}[t]{0.31\textwidth}
		\includegraphics[width=\textwidth]{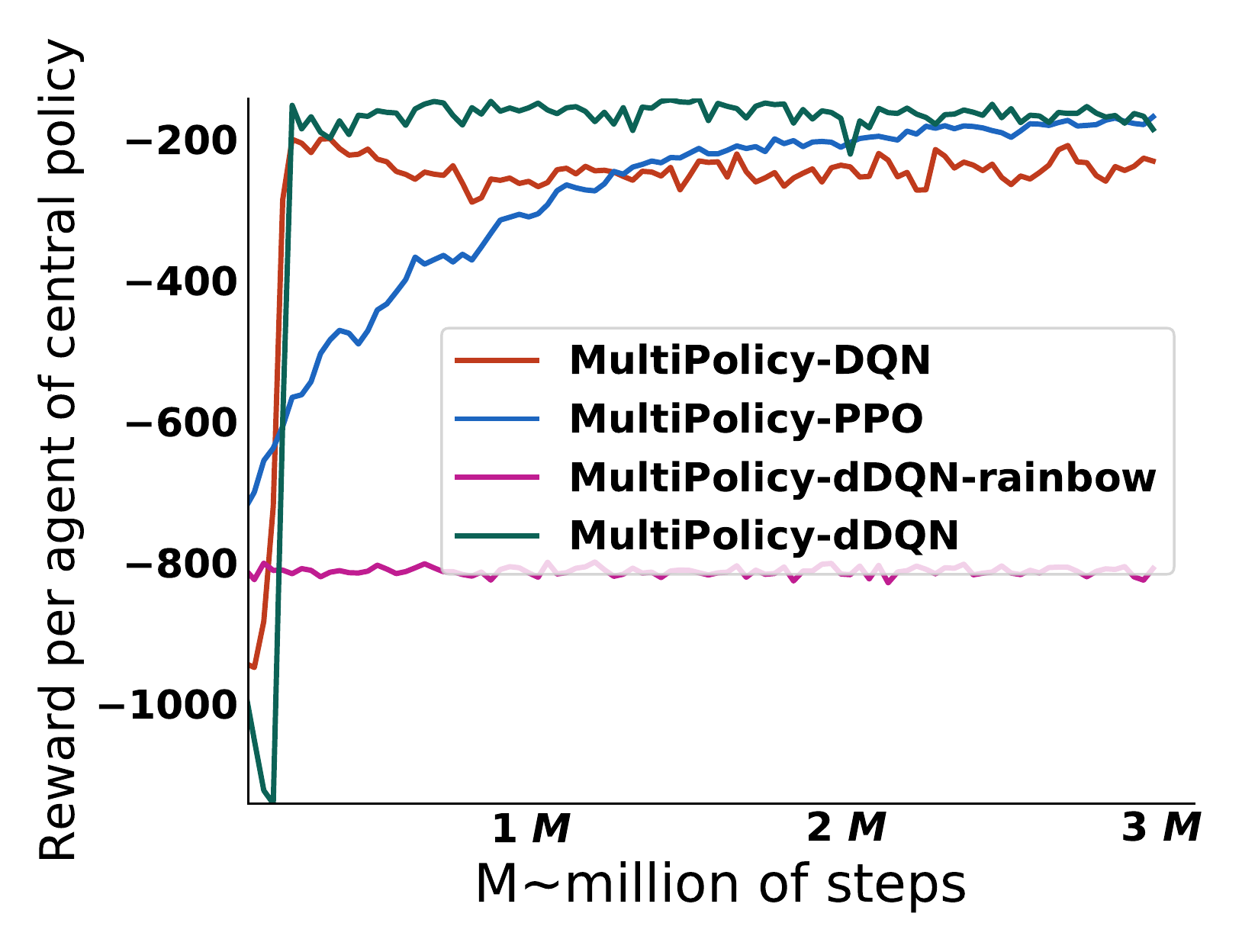}
		\caption{Average reward per agent for central policy~(reward defined in Eq.~(\ref{eq:central-reward})).}
		\label{fig:policy-central}
	\end{subfigure}
	\begin{subfigure}[t]{0.31\textwidth}
		\includegraphics[width=\textwidth]{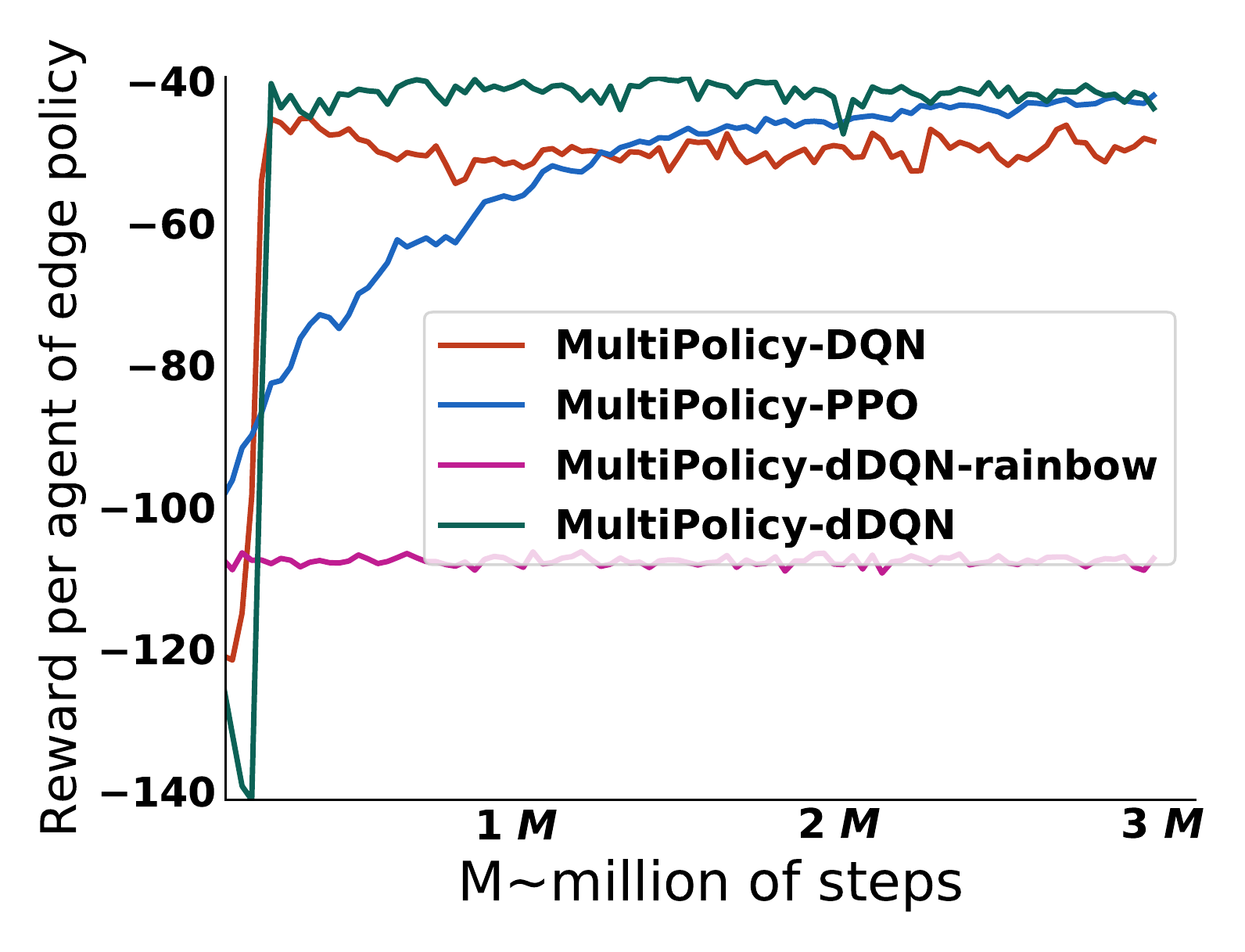}
		\caption{Average reward per agent for edge policy~(reward defined in Eq.~(\ref{eq:edge-reward})).}
		\label{fig:policy-edge}
	\end{subfigure}
	\caption{Average reward per agent.}
	\label{fig:policy}
\end{figure*}

We conduct the training over iterations, each of which consists of numerous rollouts.\footnote{Rollout, or playout, is a term often used in machine learning that is defined by Monte Carlo. In each rollout, an agent takes actions until reaching predefined max steps.} As shown in Fig.~\ref{fig:episode-shared} to Fig.~\ref{fig:policy}, the interval consists of 100 iterations, each of which consists of 30 rollouts. Each rollout has 1000 steps. A step maps to one second in real life. As such, the interval consists of 3 million steps.

Since \emph{SharedPolicy} and \emph{MultiPolicy} have different reward definitions, their reward performances are not strictly comparable. Therefore, we plot them separately. As shown in Fig.~\ref{fig:episode-shared} to Fig.~\ref{fig:policy}, \emph{dDQN} and \emph{DQN} show similar curves and the former performs slightly better than the latter. \emph{PPO} is always outperformed by \emph{dDQN} and \emph{DQN} in the beginning of the training and climbs to a similar level after around 2M steps. \emph{Rainbow} setup provides the worst performance out of all algorithms. It indicates that the values of the parameters discovered in \emph{Rainbow} does not fit our use case. Single-agent algorithms take a lot longer to reach a similar level of performance, thus we exclude their curves from the figures. For instance, the training time of single-agent \emph{DQN} is 4 times the training time of its multi-agent counterpart. This is because the action space dimension for each agent in multi-agent algorithms is $2$ while single-agent \emph{DQN} has an action space of $2^{25}$, and thus requires a much longer training time.

\subsection{Traffic Results}
\label{subsec:traffic}
We replay the policies with \texttt{SUMO} to compare the traffic control performances and show the summarized results in Table~\ref{tab:results}. 
\textit{Halting vehicles} refer to the average number of the  halting vehicles per step~(speed $<$ 0.1\,m/s). \textit{Queuing time} indicates the average waiting time of vehicles due to  queuing per step. \textit{Queuing length} refers to the average length of the queues per step. The end of the queue is defined as the last vehicle with a speed of $<$ 5\,km/h. \textit{Speed} refers to the average speed. 

\begin{figure*}[!tb]
\centering
\captionsetup{justification=centering}
	\begin{subfigure}[t]{0.31\textwidth}
		\includegraphics[width=\textwidth]{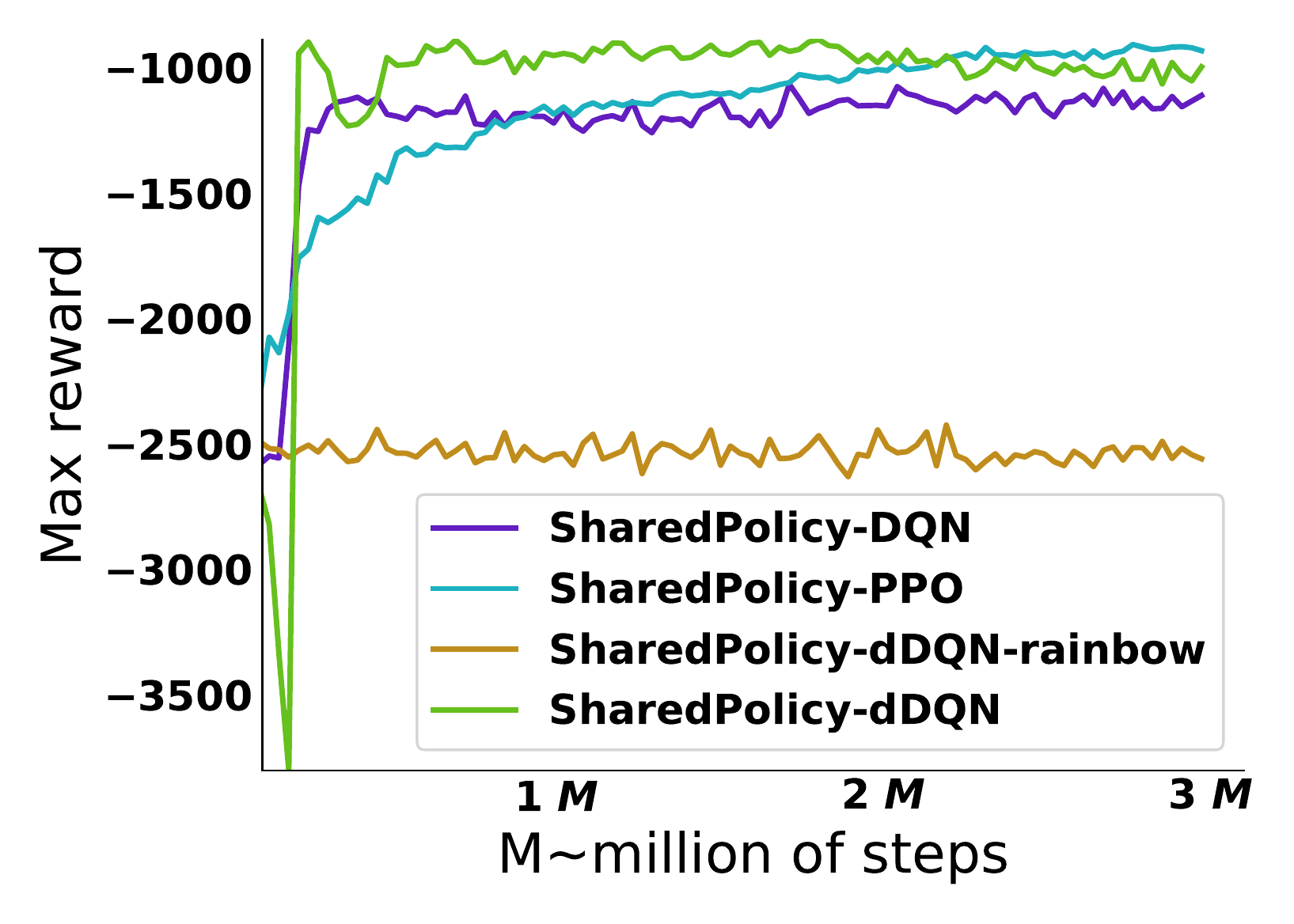}
		\caption{Max reward.}
		\label{fig:max-shared}
	\end{subfigure}
	\begin{subfigure}[t]{0.31\textwidth}
		\includegraphics[width=\textwidth]{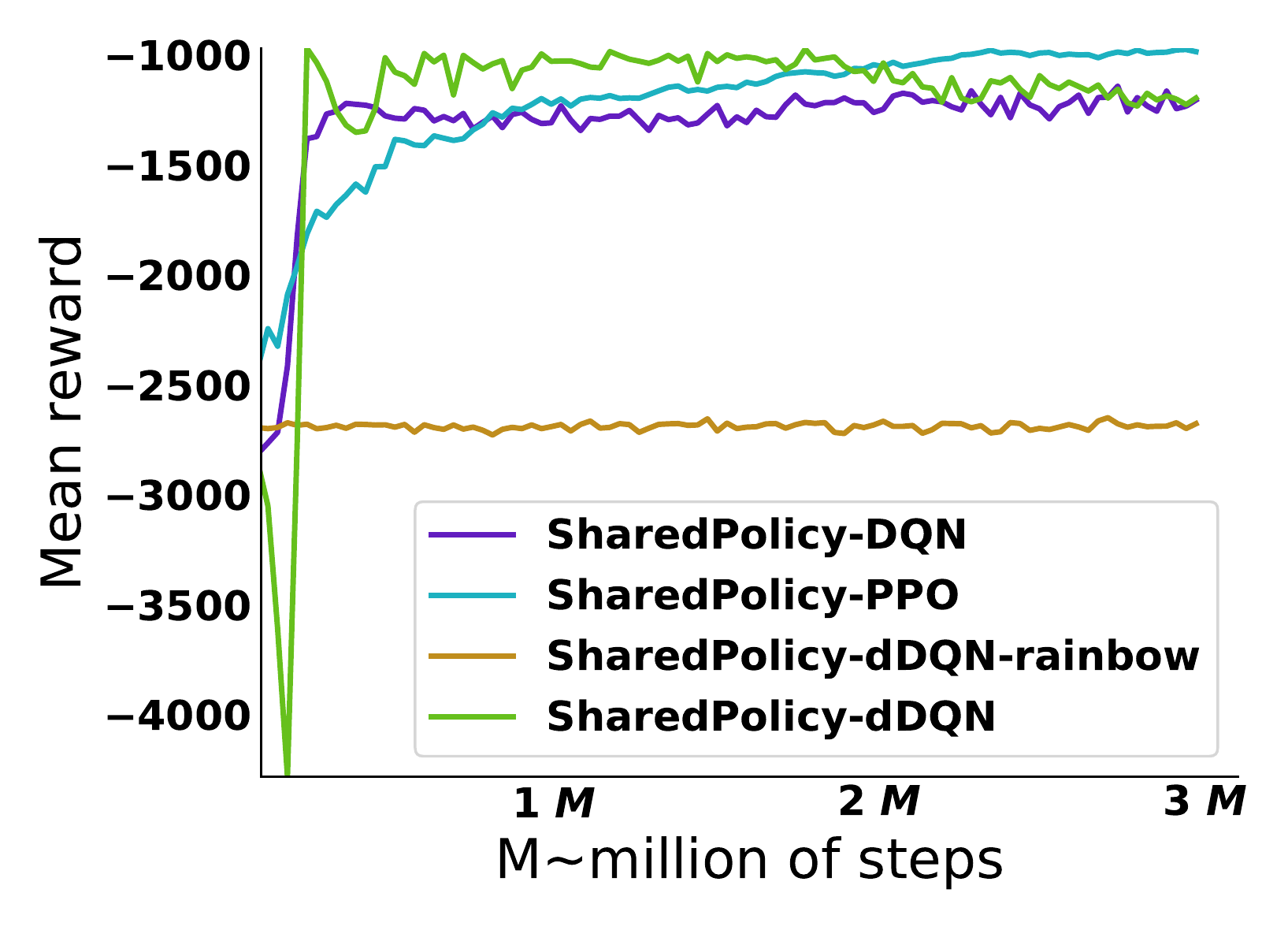}
		\caption{Average reward.}
		\label{fig:mean-shared}
	\end{subfigure}
	\begin{subfigure}[t]{0.31\textwidth}
		\includegraphics[width=\textwidth]{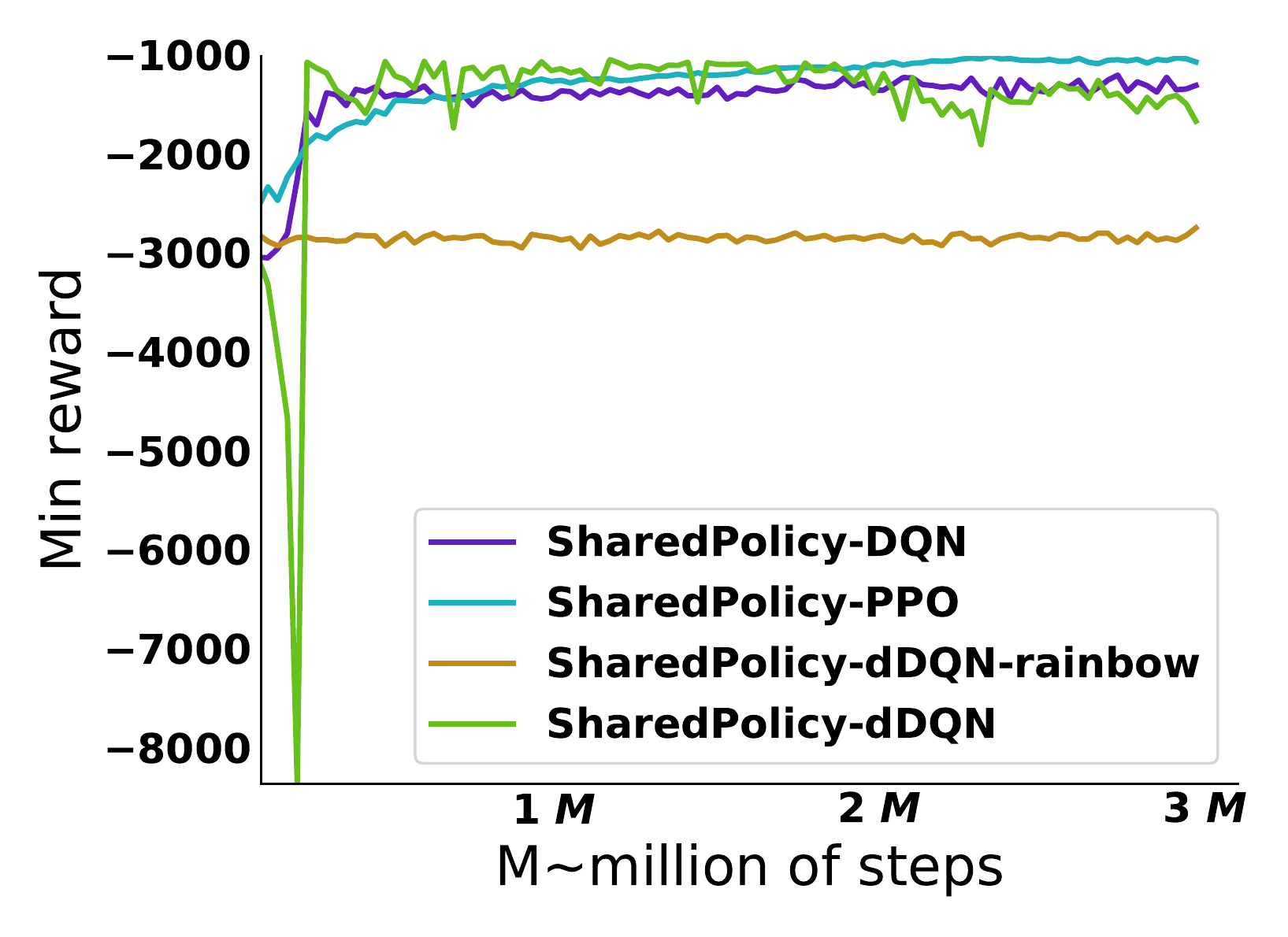}
		\caption{Minimum reward.}
		\label{fig:min-shared}
	\end{subfigure}
	\caption{Reward for \emph{SharedPolicy} (defined in Eq.~(\ref{eq:share-reward})).}
	\label{fig:episode-shared}
\end{figure*}
\begin{figure*}[!tb]
\centering
\captionsetup{justification=centering}
	\begin{subfigure}[t]{0.31\textwidth}
		\includegraphics[width=\textwidth]{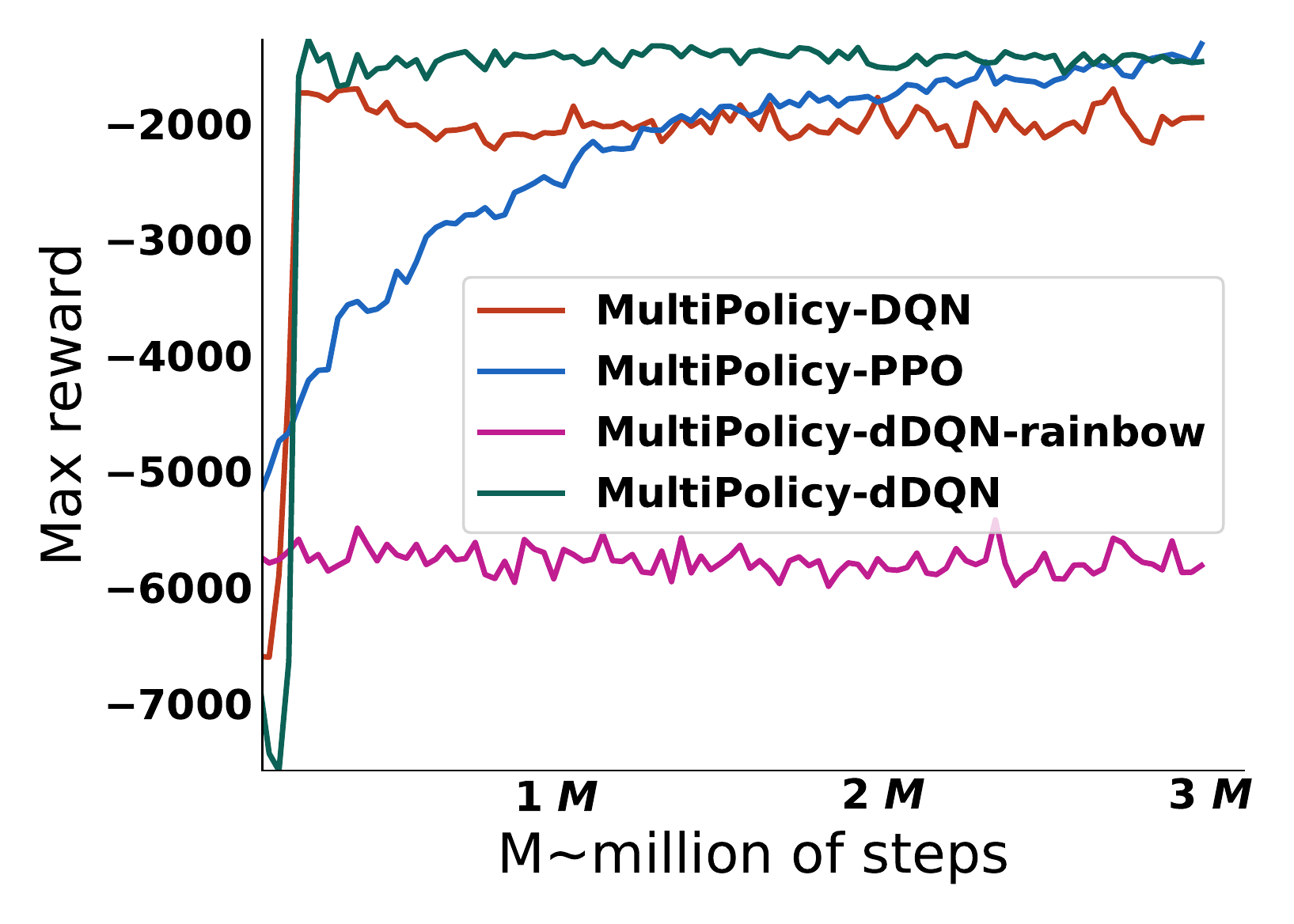}
		\caption{Max reward.}
		\label{fig:max-multi}
	\end{subfigure}
	\begin{subfigure}[t]{0.31\textwidth}
		\includegraphics[width=\textwidth]{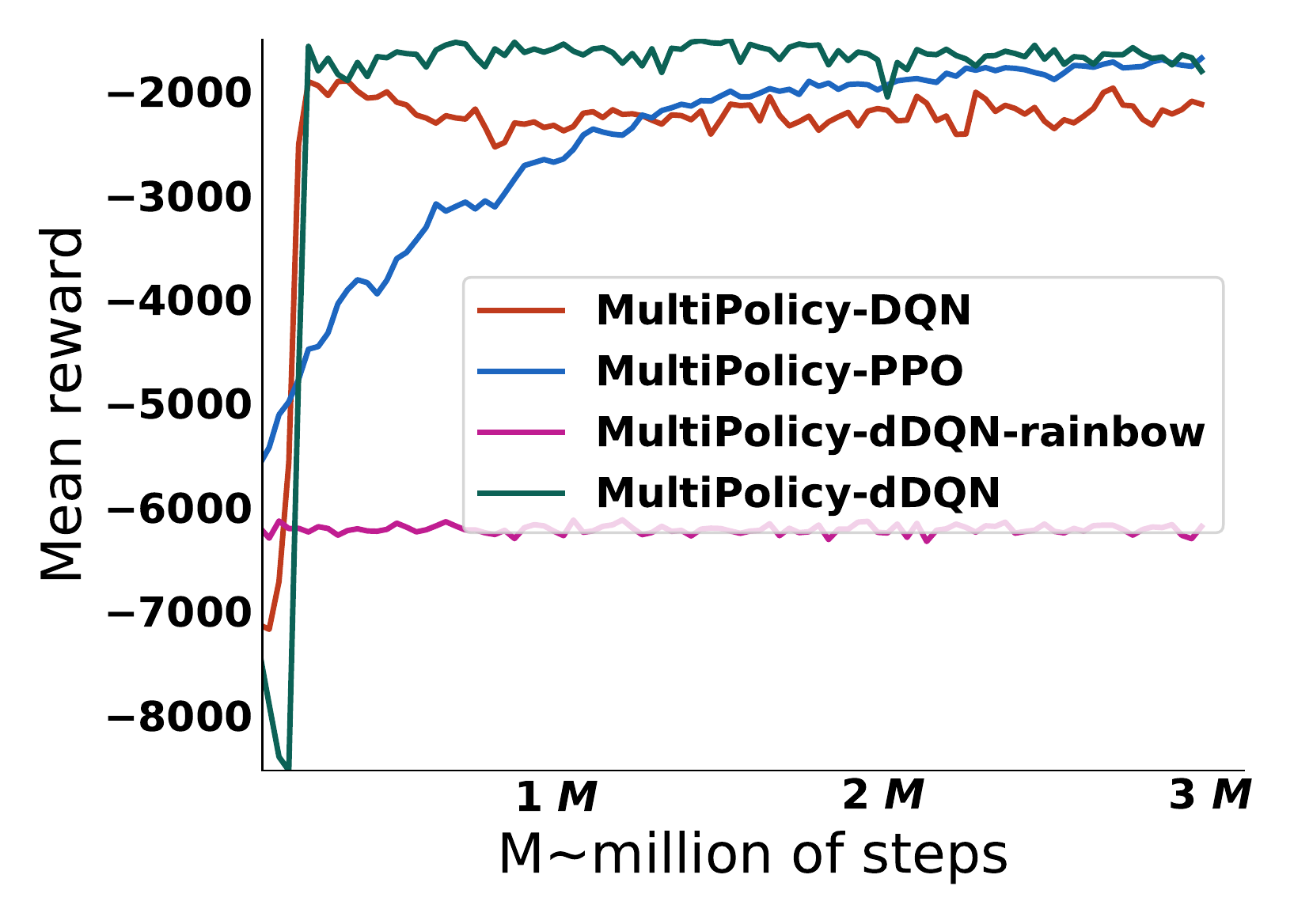}
		\caption{Average reward.}
		\label{fig:mean-multi}
	\end{subfigure}
	\begin{subfigure}[t]{0.31\textwidth}
		\includegraphics[width=\textwidth]{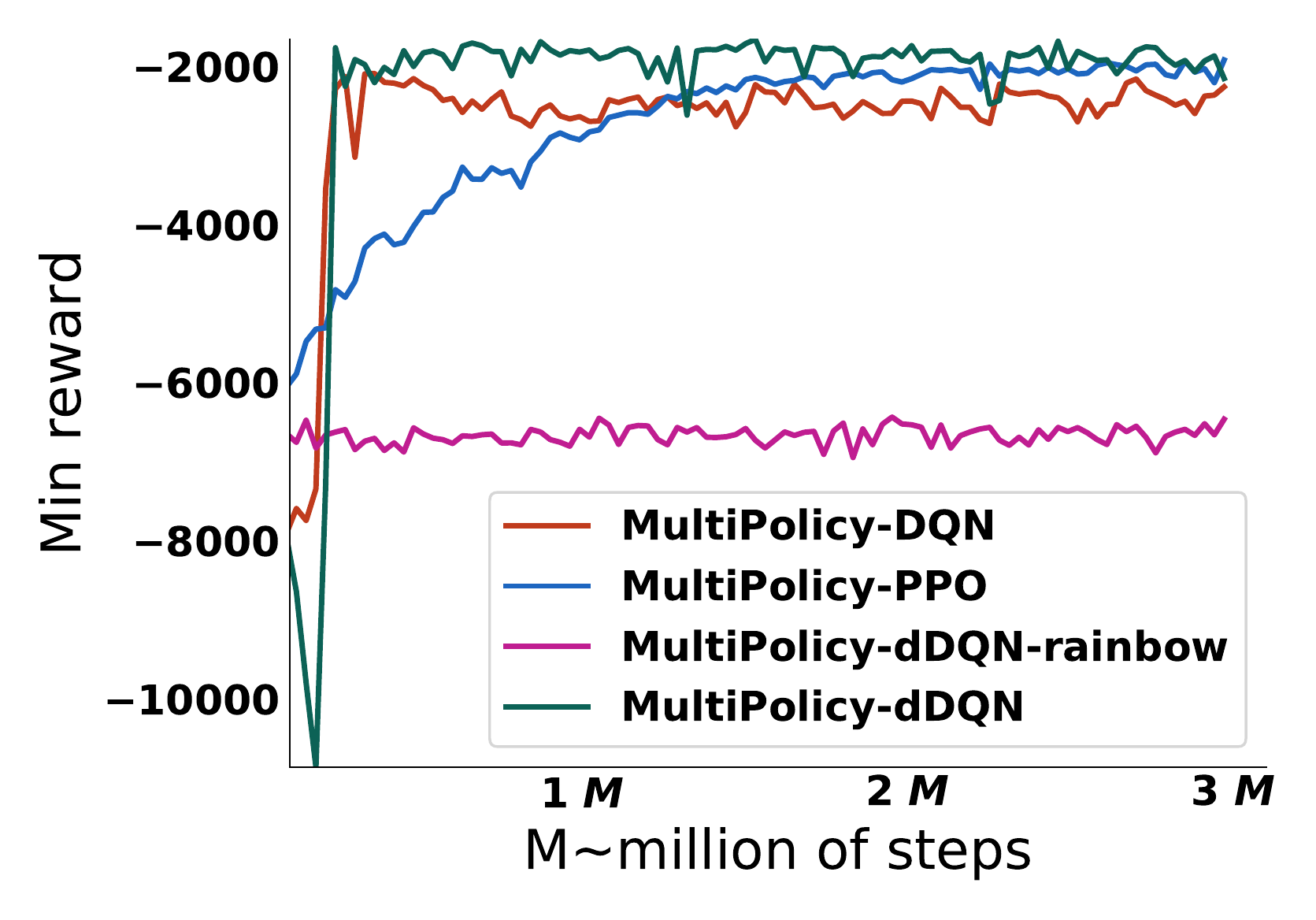}
		\caption{Minimum reward.}
		\label{fig:min-multi}
	\end{subfigure}
	\caption{Reward for \emph{MultiPolicy} (defined in Eq.~(\ref{eq:central-reward}) and Eq.~(\ref{eq:edge-reward})).}
	\label{fig:episode-multi}
\end{figure*}
As shown, the \emph{Static} algorithm presents the worst performance. Meanwhile, we find that all machine learning algorithms provide very limited improvement for the average vehicle speed compared to the \emph{Actuated} algorithm. However, they all decrease the number of halting vehicles and queuing time significantly (except \emph{Rainbow} algorithms which is most likely due to unfeasible parameter values). It indicates that with the agent-controlled signals, vehicles have much less start/stop waves and waiting time in queues, thus gaining considerable improvements in driving experience.

\renewcommand{\arraystretch}{1.5}
\begin{table*}[t]
\centering
\caption{Traffic performance and required training time. \\N: Non-ML, S: SingleAgent-ML, M: Multiagent-ML}
\label{tab:results}
\begin{tabular}{;c|^l^r^r^r^r}
\specialrule{1.3pt}{2pt}{2pt}

Num\_agents & Algorithm & Halting vehicles & Queue time (s) & Queue length (m) & Speed (m/s) \\
\midrule

\multirow{ 2}{*}{N} 
&     \emph{Static} & 95 \textpm 40 &  14.23 \textpm 10.25 & 19.82 \textpm 12.17   &13.39 \textpm 3.20  \\ \cline{2-6}
& \emph{Actuated} &13.14 \textpm 6.63& 3.58 \textpm 4.65 & 14.85 \textpm 10.57  &18.25 \textpm 1.96 \\ \cline{1-6}
\multirow{ 4}{*}{S}
&      \emph{ARS} & 3.34 \textpm 3.44& 0.27 \textpm 0.44 & 7.18 \textpm 0.99 & 16.39 \textpm 1.56  \\  \cline{2-6} 
&      \emph{ES} &2.87 \textpm 3.11& 0.30 \textpm 0.51 & 7.31 \textpm 1.29  & 17.22 \textpm 1.67  \\  \cline{2-6} 
&     \emph{PPO} & 9.78 \textpm 3.51& 0.76 \textpm 1.26 &8.20 \textpm 3.38 & 15.06 \textpm 1.52 \\ \cline{2-6} 
&      \emph{DQN} & 2.06 \textpm 1.78 & 0.25 \textpm 0.50 & 7.31 \textpm 1.29  & 16.67 \textpm 1.56  \\ \cline{2-6}
\cline{1-6}

\multirow{6}{*}{M}
&   \emph{SharedPolicy-PPO} &1.74 \textpm 1.48 & 0.23 \textpm 0.00 & 7.32 \textpm 0.24 & 18.19 \textpm 0.47 \\ \cline{2-6} 
&   \emph{SharedPolicy-DQN} & 2.24 \textpm 1.48 & 0.27 \textpm 0.00 & 7.44 \textpm 0.21 & 18.06 \textpm 0.44   \\  \cline{2-6} 
& \emph{SharedPolicy-dDQN-rainbow} & 11.58 \textpm 4.45 & 0.88 \textpm 0.00 & 8.63 \textpm 1.16 & 16.17 \textpm 0.55  \\ \cline{2-6} 
&   \emph{SharedPolicy-dDQN} & 2.38 \textpm 1.48 & 0.68 \textpm 0.00 & 9.23 \textpm 0.27 & 18.03 \textpm 0.79   \\ \cline{2-6} 
&   \emph{MultiPolicy-PPO}  & 3.79 \textpm 2.97 & 0.30 \textpm 0.00 & 7.19 \textpm 0.16 & 17.60 \textpm 0.61 \\ \cline{2-6} 
&    \emph{MultiPolicy-DQN} & 2.13 \textpm 1.48 & 0.26 \textpm 0.00 & 7.34 \textpm 0.21 & 17.97 \textpm 0.53   \\ \cline{2-6} 
&   \emph{MultiPolicy-dDQN-rainbow} & 11.81 \textpm 2.97 & 0.88 \textpm 0.00 & 8.88 \textpm 1.16 & 16.08 \textpm 0.46  \\ \cline{2-6} 
&   \emph{MultiPolicy-dDQN} & 3.74 \textpm 2.97 & 0.37 \textpm 0.00 & 7.31 \textpm 0.16 & 17.65 \textpm 0.55  \\  
\specialrule{1.3pt}{1pt}{1pt}
\end{tabular}
\end{table*}

Multi-agent \emph{SharedPolicy-PPO}, \emph{SharedPolicy-DQN} and \emph{MultiPolicy-DQN} provide the smoothest traffic control. Their performances show the smallest numbers of halting vehicles and shortest queuing time and lengths while allowing higher driving speeds than others. The performance of single-agent \emph{DQN} is fairly good. However it requires exponentially higher capacity and four times as long as the training period of its multi-agent counterparts. \emph{ARS} and \emph{ES} also show competitive performances, although presenting either slightly slower speeds or more halting vehicles. Nevertheless, \emph{ARS} and \emph{ES} currently only support single-agent training, thus also require much longer training time. For example, to achieve the performance in Table~\ref{tab:results}, \emph{ARS} and \emph{ES} need around 800K steps~(approximately 253 minutes) while multi-agent \emph{DQN} and \emph{dDQN} only need around 240K steps~(approximately 52 minutes). Similarly, \emph{PPO} has much longer convergence time compared to multi-agent \emph{DQN} algorithms as shown in Fig.~\ref{fig:episode-shared} and Fig.~\ref{fig:episode-multi}.

Overall, the training and traffic results align with Theorem 1 and 2 proposed in~\cref{sec:algorithm}, showing that decentralized multi-agent \emph{DQN} can achieve similar optima with centralized single-agent \emph{DQN}. Surprisingly, \emph{MultiPolicy} has not provided observable benefits compared to \emph{SharedPolicy}. We will explore more varieties of \emph{MultiPolicy} setups in future works. 
Meanwhile, decentralized multi-agent \emph{DQN} outperforms centralized single-agent \emph{DQN} in the following aspects:
\begin{enumerate*}
    \item It requires only 1/4 of the training time centralized \emph{DQN} needs.
    \item It requires exponentially less computation and memory capacity than centralized \emph{DQN}.
\end{enumerate*}

Next, we evaluate the scalability of~\sysname by conducting tests on different scales of maps. Thereafter we evaluate the communication overhead with a preliminary simulation.

\begin{figure}
		\centering
		\includegraphics[width=.7\linewidth]{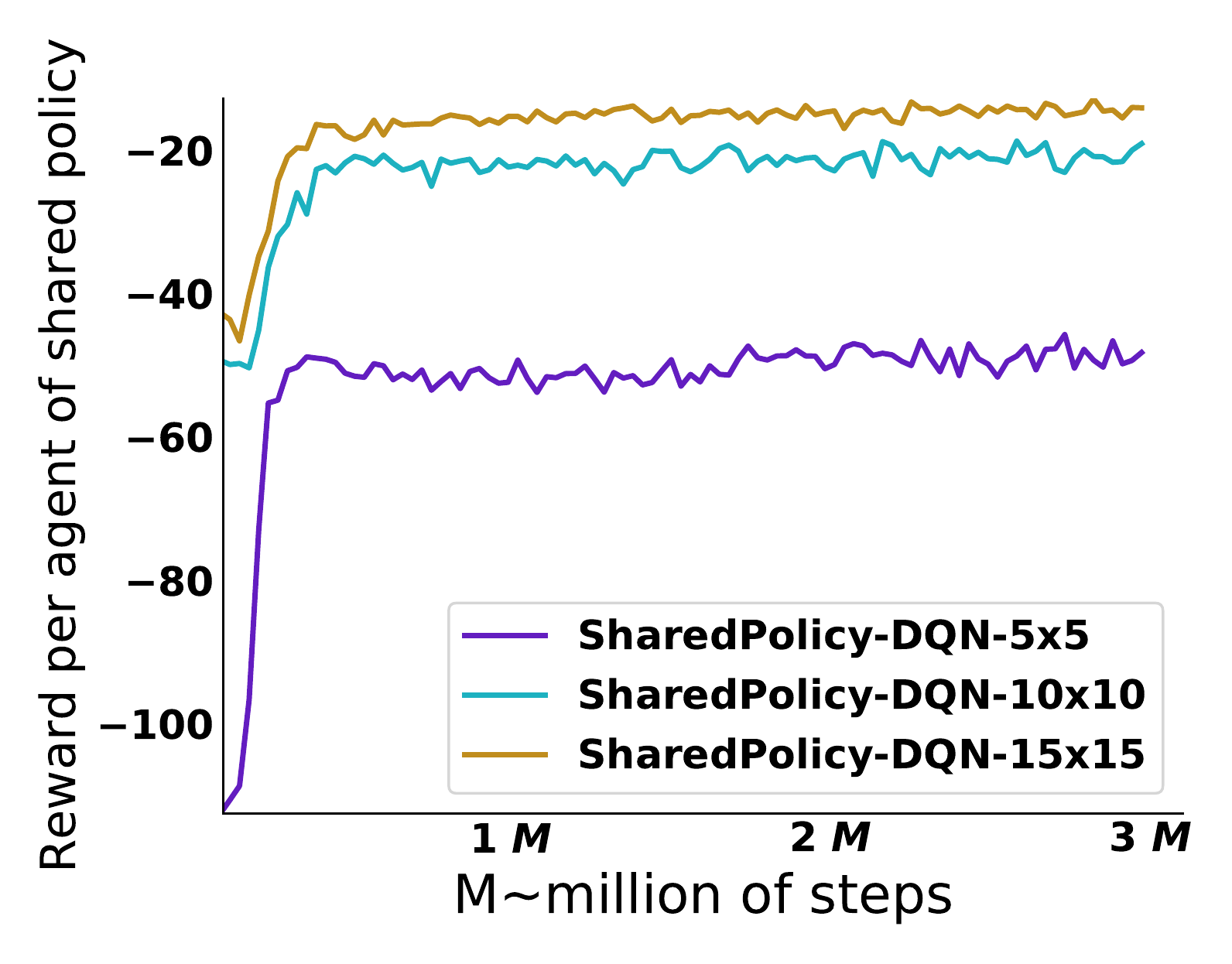}
		\caption{Average reward per agent for \emph{SharedPolicy-DQN} on different scales.}
		\label{fig:scale}
\end{figure}
\renewcommand{\arraystretch}{1.5}
\begin{table*}[t]
\centering
\caption{Performance of \emph{SharedPolicy-dDQN} on different map scales.}
\label{tab:traffic-scale}
\begin{tabular}{c|^c^c^c^c|c}
\specialrule{1.3pt}{2pt}{2pt}
Scale & Halting vehicles & Queue time (s) & Queue length (m) & Speed (m/s) & Training time/rollout (s) \\
\midrule
$5\times5$ & 2.24 \textpm 1.48 & 0.27 \textpm 0.00 & 7.44 \textpm 0.21 & 18.06 \textpm 0.44 &10.22 \\\hline
$10\times10$ &7.54 \textpm 2.97 & 0.24 \textpm 0.00 & 7.27 \textpm 0.19 & 17.21 \textpm 0.30  & 35.50 \\ \hline
$15\times15$ & 13.61 \textpm 5.93 & 0.22 \textpm 0.00 & 7.25 \textpm 0.22 & 16.91 \textpm 0.22 &86.31\\
\specialrule{1.3pt}{1pt}{1pt}
\end{tabular}
\end{table*}
\subsection{Scalability}

To illustrate the system scalability, we apply multi-agent \emph{SharedPolicy-DQN} on various map scales. Fig.~\ref{fig:scale} shows the training performances of \emph{SharedPolicy-DQN} on maps with $5\times5$, $10\times10$, and $15\times15$ intersections. Because of the fixed vehicle inflow rate~(see \cref{subsec:setup}), each intersection has less nearby vehicles in larger maps, resulting in shorter queue lengths and vehicle delay, which, in turn, leads to a higher reward. As such, the average reward per agent increases as the map scale increases. The traffic control performance summarized in Table~\ref{tab:traffic-scale} mostly confirms this reasoning. 
It is important to note that when we increase the size of the map, the number of halting vehicles per intersection decreases. Overall, \sysname performs well on different map scales.

\subsection{Communication Overhead}
\label{subsec:comm}

~\sysname leverages the ubiquity of the IoV and the foreseeable edge facilities in the near future to run the algorithms. In this subsection, we provide a preliminary evaluation of the communication overhead brought by this system architecture.

We deploy a map consisting of $5\times5$ intersections as shown in Fig.~\ref{fig:map} in \texttt{NS-3}, a packet-level discrete-event network simulator for internet systems. Because the mobility trace files generated from the training tests are not compatible with \texttt{NS-3}, we use the trace files instead to estimate the average number of active vehicles per step in the considered map, i.e., 230. Then we simulate 230 vehicles in \texttt{NS-3} with the same parameters (speed and acceleration etc.) with the training tests. As such, we argue that the transmission delay performance, albeit not identical to the performance in an integrated test, should be very similar from a statistical perspective. As described in Subsection~\ref{subsec:comm_analysis}, each vehicle sends 1500\,B-packets to a tri-sector eNB at the center of the map at 1\,Hz via C-V2X~(LTE)~(\cref{tab:comm-metric}). For a period of 1000 seconds (the default training period in training tests), the transmission delay result shows that uplink (vehicle-to-$\text{ES}_1$) plus downlink ($\text{ES}_1$-to-signal) delay is less than 240\,ms per packet~(Table~\ref{tab:transdelay}). As such, the transmission delay during each step is much smaller than the step duration\footnote{See footnote 2.} thus does not impact on~\sysname.

\renewcommand{\arraystretch}{1.4}
\begin{table}[t]
\centering

 \centering
 \caption{Communication parameters.}
 \label{tab:comm-metric}
 \begin{tabular}{;c^c^c^c^c}
    \specialrule{1.3pt}{1pt}{1pt}
    Model & Protocol & Packet & Frequency & Hops \\ 
    \midrule
    V2N~\cite{3gppv2n} & UDP & 1500 B & 1\,Hz &  1  \\
    \specialrule{1.3pt}{1pt}{1pt}
\end{tabular}

\end{table}
\begin{table}[t]
\centering

 \centering
 \caption{Transmission delays. (MAD: Median Absolute Deviation)}
 \vspace{1mm}
 \label{tab:transdelay}
 \begin{tabular}{;c|^c^c}
    \specialrule{1.3pt}{1pt}{1pt}
    Direction & Mean (ms)  & MAD (ms)  \\ 
    \midrule
    Uplink (vehicle-to-$\text{ES}_1$) & 110.82 & 17.68  \\\hline
    Downlink ($\text{ES}_1$-to-signal) & 106.23 & 0  \\
    \specialrule{1.3pt}{1pt}{1pt}
\end{tabular}

\end{table}

\section{Conclusion and Future work}
\label{sec:conclusion}
In this work, we present \sysname, an integrated edge computing framework leveraging the ubiquity of the IoV for alleviating traffic congestion in real time at city scale. We decompose the highly complex centralized problem of large area traffic light control to a multi-agent decentralized problem and prove its global optima with concrete mathematical reasoning. \sysname exploits the low latency of edge servers to provide fast \emph{DQN} training and control feedback. Thanks to its layered architecture and hierarchical algorithm, \sysname runs optimization at the \textit{Intersection}, \textit{Intra-ES} and \textit{Inter-ES} levels that allows for traffic light control on different scales. We present numerous comparisons to evaluate the traffic improvement brought by~\sysname and show that, compared to the state-of-the-art baseline works, \sysname decreases the convergence time by 65.66\% compared to \emph{PPO} and training steps by 79.44\% compared to \emph{ARS} and \emph{ES}. Besides, \sysname provides comparable traffic control performance with its centralized counterpart while requiring only 1/4 of the training time.

This paper explores the performance of algorithm on \textit{Intra-ES} level. In future work, we will explore the \textit{Inter-ES} algorithm and its impact on the performance of~\sysname. We are also looking forward to investigating the potential of tuning the threshold parameter value in the \textit{Intersection} level algorithm instead of directly switching the traffic lights, to further decrease the action space and extend the scalability.

\end{document}